\documentclass[prd,aps,floatfix,nofootinbib, 10 pt]{revtex4}

\newcommand{\bea}{\begin{eqnarray}}
\newcommand{\eea}{\end{eqnarray}}
\newcommand{\beq}{\begin{eqnarray}}
\newcommand{\eeq}{\end{eqnarray}}
\newcommand{\nn}{\nonumber}

\usepackage{amsmath,graphicx,color,epsfig}
\begin{document}
\title{Analysis of Forward-Backward and Lepton Polarization Asymmetries in $B\to K_{1}\ell^{+}\ell^{-}$ Decays in the Two-Higgs-doublet Model}
\author{Naveed Ahmed$^{1}$}
\author{, Ishtiaq Ahmed$^{2,3}$}
\author{, M. Jamil Aslam$^{4}$}
\affiliation{$^{1}$Department of Physics,
Federal Urdu University of Arts, Science and technology, Islamabad 45320, Pakistan}
\affiliation{$^{2}$National Centre for Physics,
Quaid-i-Azam University Campus, Islamabad 45320, Pakistan }\affiliation{$^3$Laboratorio de F\'{\i}sica Te\'{o}rica e Computa\c{c}\H{a}o Científica,
Universidade Cruzeiro do Sul, 01506-000, S\H{a}o Paulo, Brazil}\affiliation{$^4$Department of Physics,
Quaid-i-Azam University Campus, Islamabad 45320, Pakistan }

\begin{abstract}
The exclusive semileptonic $B\to K_{1}(1270) \ell^{+}\ell^{-}$ ($\ell=\mu , \tau$) decays are analyzed in variants of two Higgs double models (THDMs). The mass eigenstates $K_{1}(1270)$ and $K_{1}(1400)$ are the mixture of two axial-vector SU(3)  $^{1}{P}_{1}$ and $^{3}{P}_{1}$ states with the mixing angle $\theta_{K}$. Making use of the form factors calculated in the Light Cone QCD approach and by taking the mixing angle $\theta_{K}=-34^{\circ}$, the impact of the parameters of the THDMs on different asymmetries in above mentioned semileptonic $B$ meson decays are studied. In this context the forward-backward asymmetry and different lepton polarization asymmetries have been analyzed. We have found comprehensive effects of the parameters of the THDMs on the above mentioned asymmetries. Therefore, the precise measurements of these asymmetries at the LHC and different $B$ factories, for the above mentioned processes, can serve as a good tool to put some indirect constraints on the parametric space of the different versions of THDM.
\end{abstract}

\maketitle

\section{Introduction}\label{intro}

The Standard Model (SM) of particle physics successfully explain the observed data so far and the recent observation of a Higgs (like) boson with the mass range of 126 GeV support it further. However, it is still far to believe as an ultimate theory of nature. The LHC data is now ready to have further analysis which possibly check the SM in more detail and also probe down to New Physics (NP). It is, therefore, an exciting time to test the predictions of the SM in different sector and try to identify the nature of physics beyond it. The study of the rare decays of $B$ mesons induced by the flavor changing neutral current transitions (FCNC), being loop suppressed in the SM,  provides us a natural ground to look for the possible existence of the NP at TeV scale associated with the hierarchy problem.

The measurements of the inclusive $b \to s \ell^{+} \ell^{-}$ transitions are preferred because of the lower theoretical uncertainties. However, they are most challenging to measure experimentally. The branching fractions and various asymmetries of the inclusive $B \to X_s \ell^{+}\ell^{-}$ decays, where $\ell$ can be any of the three leptons and $X_{s}$ is any hadronic state with $s$ quark are measured at Belle \cite{Belle} and \textit{BABAR} \cite{Babar}. The theoretical studies of the rare $B$ meson decays give an opportunity to investigate the physics beyond SM, where these decays are purposefully used to test these models and to constrain the parameter space of these models.

It is remarkable that most of the experimental results are in agreement with the SM predictions but the problems such as neutrino oscillations, the matter-antimatter asymmetry and the problems of dark matter can not be explained in this model. It is, therefore, widely believed that it is an effective theory at an electroweak scale. In order to understand these unfinished mysteries of nature, there exist some physics which lies beyond the scope of the SM or need its extensions. In this context, some possible extensions of the SM including the little Higgs model \cite{fal3, fal4},  the extra-dimension model \cite{fal5, fal6}, and multi-Higgs models like the supersymmetric standard model \cite{hati7} are extensively studied in literature.

Two Higgs doublet model (THDM) is among the most popular extensions of the SM. Contrary to the SM in where we have only one Higgs doublet, in the THDM we consider two complex Higgs doublets. Generally, the THDM posses tree level FCNC transitions which can be avoided by imposing an ad-hoc discrete symmetry \cite{adhoc}. This results to two different possibilities:

\begin{itemize}
\item The first possibility to keep the flavor conservation at the tree level is to couple all the fermions to only one of the Higgs doublet. It is called to be the model I.
\item The second possibility is when the up-type quarks are coupled to the one Higgs double and the down-type to the second one and it is named as model II. This is a popular choice because in this case the Higgs sector coincides with the supersymmetric model.
\end{itemize}
The physical contents of the Higgs sector contain two neutral scalar Higgs bosons $H^{0}$, $h^{0}$, a psudo-scalar Higgs $A^{0}$ and pair of the charged Higgs boson $H^{\pm}$. The vacuum expectation values of the two Higgs doublets are denoted by $v_{1}$ and $v_{2}$, respectively, and the interaction of the fermions to the Higgs fields depend on the $tan\beta = v_{2}/v_{1}$ which is  a free parameter of the model.

There is another possibility where the discrete symmetry is not imposed  which in turn leads to the most general form of the THDM, i.e., to say model III. In this version the FCNC transitions are allowed at the tree level. The indirect constraints on the masses of charged Higgs bosons $m_{H^{\pm}}$, the neutral scalars $m_{H^0}, m_{h^0}$, the neutral psudo-scalar $m_{A^0}$  as well as on the fermion Higgs interaction vertex, $tan{\beta}$ are obtained from the experimental observation of branching ratios of $b \to s \gamma$, $B \to D \tau \nu_{\tau}$ decays  and $K - \bar{K}$ and $B - \bar{B}$ mixing in the literature \cite{constraints}.  Consistent with the low energy constraints, the FCNCs involving the third generation are not as severely suppressed as the one involving the first two generations. In contrast to the SM and THDMs I and II there exist a single $CP$ phase of vacuum which leads to a rich source of the phenomenological studies of $CP$ violating observables \cite{Falahati}.

In connection with the FCNC transitions mediated by $b \to s \ell^{+} \ell^{-}$, like the rare semileptonic decays involving $B\rightarrow (X_{s},K^{\ast},K)\ell^{+}\ell^{-}$, the $B\rightarrow K_{1}(1270,1400)\ell^{+}\ell^{-}$ decays are also rich in phenomenology to get some hints of the NP \cite{IPJ}. In some sense they might be even more interesting and sophisticated to NP because of the mixture of $K_{1A}$ and $K_{1B}$, where $K_{1A}$ and $K_{1B}$ are the members of two axial-vector SU(3) octect $^3P_1$ and $^1P_1$ states, respectively. The physical states $K_{1}(1270)$ and $K_{1}(1400)$ can be obtained by the mixing of $K_{1A}$ and $K_{1B}$ as
\begin{subequations}
\begin{eqnarray}
\vert K_{1}(1270)\rangle &=& \vert K_{1A}\rangle\sin{\theta_{K}}+\vert K_{1B}\rangle\cos{\theta_{K}},\label{mix1}\\
\vert K_{1}(1400)\rangle &=& \vert K_{1A}\rangle\cos{\theta_{K}}-\vert K_{1B}\rangle\sin{\theta_{K}},\label{mix2}
\end{eqnarray}
\end{subequations}
where the magnitude of mixing angel $\theta_{K}$ has been estimated to be $34^{\circ}\leq |\theta_{K}|\leq 58^{\circ}$  \cite{Suzuki}. Recently, from the studies of $B\to K_{1}(1270)\gamma$ and $\tau \to K_{1}(1270)\nu_{\tau}$, the value of $\theta_{K}$ has been estimated to be $\theta_{K}=-(34\pm13)^{\circ}$, where the minus sign of $\theta_{K}$ is related to the chosen phase of $\vert K_{1A}\rangle$ and $\vert K_{1B}\rangle$ \cite{HY}. Getting an independent conformation of this value of mixing angle $\theta_{K}$ is by itself interesting. It has already been pointed out that this particular choice suppresses the ${\cal BR}$ for $K_{1}(1400)$ in the final state compared to $K_{1}(1270)$, which can be tested in at some on going and future experiments \cite{aqeel}.

There exists extensive studies showing that the observables such as branching ratio $(\mathcal{BR})$, the forward-backward asymmetry $(\mathcal{A}_{FB})$, lepton polarization asymmetries $(P_{i})$ and helicity fractions of the final state meson $f_{L,T}$ for semileptonic $B$ decays are greatly influenced in different beyond the SM scenarios \cite{IPJ}. Therefore, the precise measurement of these observables will play an important role in the indirect searches of NP and possibly the signatures of the THDM. The purpose of present study is to addresses this question i.e., to investigate the possibility of searching NP due to variants of THDMs  in $B\rightarrow K_{1}(1270, 1400)\ell^{+}\ell^{-}$ decays with $\ell=\mu,\tau$ through forward-backward asymmetry and lepton polarization asymmetry.

The manifestations of the NP due to the THDM is two fold in a sense that it modifies the Wilson coefficients as well as it introduces the new operators in the effective Hamiltonian in addition to the SM operators. In the present study, the NP effects are analyzed by studying the forward-backward asymmetry ${\cal A}_{FB}$ and the lepton polarization asymmetries  for $B \rightarrow K_{1}(1270)\ell^{+}\ell^{-}$ decays in all the three variants of THDMs, namely, models I, II and III.

The plan of the study is as follows: In sec. \ref{tf}, we fill our toolbox with the theoretical framework needed to study the said process in the THDM. In Sec. \ref{mix}, we present the mixing of $K_{1}(1270)$ and $K_{1}(1400)$ and the form factors used in this study. In Sec. \ref{obs}, we discuss the observables of $B\rightarrow K_{1}\ell^{+}\ell^{-}$ in detail, whereas, in Sec. \ref{num} we give the numerical analysis of our observables and discuss the sensitivity of these observables with the THDM parameters and finally we conclude the findings of present study. 

\section{Theoretical Framework}\label{tf}

At quark level, the semileptonic decays $B\rightarrow K_{1}(1270,1400)\ell^{+}\ell^{-}$ are governed by the transition $b\rightarrow s\ell^{+}\ell^{-}$ for which the general effective Hamiltonian in the SM and in THDM can be written, after integrating out the heavy degrees of freedom in the full theory, as \cite{Huang}:
\begin{align}
H_{eff}=-\frac{4G_{F}}{\sqrt{2}}V_{tb}V_{ts}^{\ast }&\bigg[\sum\limits_{i=1}^{10}C_{i}(\mu )O_{i}(\mu)+\sum\limits_{i=1}^{10}C_{Qi}(\mu )Q_{i}(\mu) \bigg],  \label{1}
\end{align}%
where $O_{i}(\mu )$ $(i=1,2, \cdots,10)$ are the four quark operators and $%
C_{i}(\mu )$ are the corresponding Wilson coefficients at the energy scale $%
\mu $ which is usually taken to be the $b$-quark mass $\left(
m_{b}\right) $. The theoretical uncertainties related to the renormalization
scale can be reduced when the next to leading logarithm corrections are
included. Also the contribution from the charged Higgs boson in case of the THDM is absorbed in these Wilson coefficients. The new operators $Q_{i}(i=1, 2, \cdots, 10)$ come from the NHBs exchange diagrams,
whose manifest forms and corresponding Wilson coefficients can be found in \cite{Huang} and at a scale $\mu = m_{W}$, these can be summarized as \cite{Falahati}:
\bea
C_{Q_1}(m_{W})&=&\frac{m_bm_l}{m^2_{h^0}}\frac{1}{\left|\lambda_{tt}\right|^2}\frac{1}{sin^2\theta_{W}}\frac{x}{4}\bigg[(sin^2\alpha+hcos^2\alpha)f_1(x,y) \nn \\%
  &&+\left[\frac{m^2_{h^0}}{m^2_{W}}+(sin^2\alpha+hcos^2\alpha)(1-z)\right]f_2(x,y) \notag \\
&&+\frac{sin^2 2\alpha}{2m^2_{H^{\pm}}}\left[m^2_{h^0}-\frac{(m^2_{h^0}-m^2_{H^0})^2}{2m^2_{H^0}}\right]f_3(y)\bigg] \label{cq1} \\
C_{Q_2}(m_{W})&=&-\frac{m_bm_l}{m^2_{A^0}}\frac{1}{\left|\lambda_{tt}\right|^2}{f_1(x,y)+\bigg[1+\frac{m^2_{H^\pm}-m^2_{A^0}}{2m^2_{W}}\bigg]f_2(x,y)} \label{cq2} \\
C_{Q_3}(m_{W})&=&\frac{m_{b}e^2}{m_{\ell}g^2}\bigg[C_{Q_1}(m_{W})+C_{Q_2}(m_{W})\bigg] \label{cq3} \\
C_{Q_4}(m_{W})&=&\frac{m_{b}e^2}{m_{\ell}g^2}\bigg[C_{Q_1}(m_{W})-C_{Q_2}(m_{W})\bigg] \label{cq4} \\
C_{Q_i}(m_{W})&=&0 \enskip \enskip  i= 5,...,10 \label{cq5}
\eea
where
\bea
x=\frac{m^2_t}{m^2_W}, y=\frac{m^2_t}{m^2_{H^\pm}}, z=\frac{x}{y}, h=\frac{m^2_{h^0}}{m^2_{H^0}}, \nn \\
f_1(x,y)=\frac{x\mathrm{ln}x}{x-1}-\frac{y\mathrm{ln}y}{y-1},\nn \\
 f_2(x,y)=\frac{x\mathrm{ln}y}{(z-x)(x-1)}-\frac{\mathrm{ln}z}{(z-1)(x-1)}, \nn \\
f_3(y)=\frac{1-y+y\mathrm{ln}y}{(y-1)^2}.
\eea
The evolution of the coefficients $C_{Q_1}$ and $C_{Q_2}$ is performed by the anomalous dimensions of $Q_{1}$ and $Q_2$, respectively:
\beq
C_{Q_i}(m_b)=\eta^{\gamma_Q/\beta_0}C_{Q_i}(m_W), \enskip \enskip i =1, 2
\eeq
where $\gamma_Q=-4$ is anomalous dimension of the operator $\bar{s}_Lb_R$.

The explicit forms of the operators responsible for the decay $B\rightarrow K_{1}(1270,1400)\ell^{+}\ell^{-}$, in the SM and the THDMs, are
\begin{subequations}
\begin{eqnarray}
O_{7} &=&\frac{e^{2}}{16\pi ^{2}}(m_{b}-m_s)(\bar{s}\sigma _{\mu \nu }Rb)F^{\mu
\nu }  \label{2} \\
O_{9} &=&\frac{e^{2}}{16\pi ^{2}}\left( \bar{s}\gamma _{\mu }Lb\right) \bar{\ell
}\gamma ^{\mu }\ell  \label{3} \\
O_{10} &=&\frac{e^{2}}{16\pi ^{2}}\left( \bar{s}\gamma _{\mu }Lb\right) \bar{\ell}\gamma ^{\mu }\gamma ^{5}\ell  \label{4}\\
Q_{1} &=&\frac{e^{2}}{16\pi ^{2}}\left( \bar{s}Rb\right) \bar{\ell}\ell  \label{q1}\\
Q_{2} &=&\frac{e^{2}}{16\pi ^{2}}\left( \bar{s}Rb\right) \bar{\ell}\gamma ^{5}\ell  \label{q2}\\
\end{eqnarray}%
\end{subequations}
with $L,R=\frac{1}{2}\left( 1\mp \gamma ^{5}\right)$.

Using the effective Hamiltonian given in Eq. (\ref{1}) the free quark
amplitude for $b\rightarrow s\ell ^{+}\ell ^{-}$ can be written as%
\begin{eqnarray}
\mathcal{M}(b\rightarrow s\ell^{+}\ell^{-}) &=&-\frac{G_{F}\alpha }{\sqrt{2}\pi }V_{tb}V_{ts}^{\ast }\bigg[\widetilde{C}_{9}^{eff}\left( \mu
\right) (\bar{s}\gamma _{\mu }Lb)(\bar{\ell}\gamma ^{\mu }\ell)+\widetilde{C}_{10}(\bar{s}%
\gamma _{\mu }Lb)(\bar{\ell}\gamma ^{\mu }\gamma ^{5}\ell)  \notag \\
&&-2\widetilde{C}_{7}^{eff}\left( \mu \right) \frac{m_{b}}{s}(\bar{s}i\sigma _{\mu
\nu }q^{\nu }Rb)\bar{\ell}\gamma ^{\mu }\ell+C_{Q_1}\left( \bar{s}Rb\right) \left(\bar{\ell}\ell \right) \notag\\
&&+C_{Q _2}\left( \bar{s}Rb\right) \left(\bar{\ell}\gamma^{5}\ell \right)\bigg], \label{5a}
\end{eqnarray}
where $q$ is the momentum transfer. By using the knowledge of Wilson coefficients $C_7$, $\widetilde{C}_9$ and
$\widetilde{C}_{10}$ calculated at scale $m_W$, the Wilson coefficients $\widetilde{C}_{7}^{eff}$, $\widetilde{C}_{9}^{eff}$,
$\widetilde{C}_{10}$, $C_{Q_1}$ and $C_{Q_2}$ are calculated at the scale $m_{b}$. After adding the contribution from the charged Higgs
diagrams to the SM results, the Wilson coefficients $\widetilde{C}_{7}^{eff}$, $\widetilde{C}_{9}^{eff}$ and
$\widetilde{C}_{10}$ can take the form \cite{Falahati, Huang}:
\bea
\widetilde{C}_{7}(m_W)&=&C_{7}^{SM}(m_W)+\left|\lambda_{tt}\right|^2\left(\frac{y(7-5y-8y^2)}{72(y-1)^3}+\frac{y^2(3y-2)}{12(y-1)^4}\mathrm{ln}y\right) \nn \\
&&+\lambda_{tt}\lambda_{bb}\left(\frac{y(3-5y)}{12(y-1)^2}+\frac{y(3y-2)}{6(y-1)^3}\mathrm{ln}y\right), \label{C7mw} \\
\widetilde{C}_{9}(m_W)&=&\widetilde{C}_{9}^{SM}(m_W)+\left|\lambda_{tt}\right|^2\bigg[\frac{1-4sin^2\theta_W}{sin^2\theta_W}\frac{xy}{8}\left(\frac{1}{y-1}-\frac{1}{(y-1)^2}\mathrm{ln}y\right) \nn \\
&&-y\left(\frac{47y^2-79y+38}{108(y-1)^3}-\frac{3y^3-6y^2+4}{18(y-1)^4}\right)\bigg], \label{C9mw}\\
\widetilde{C}_{10}(m_W)&=&C_{10}^{SM}(m_W)+\left|\lambda_{tt}\right|^2\frac{1}{sin^2\theta_W}\frac{xy}{8}\left(-\frac{1}{y-1}+\frac{1}{(y-1)^2}\mathrm{ln}y\right)\label{C10mw}.
\eea
It can be easily seen that in the limit $y\to 0$ along with $C_{Q_{1,2}}\to 0$ the SM results of the Wilson coefficients can be recovered.

Note that the operator $%
O_{10}$ given in Eq. (\ref{4}) can not be induced by the insertion of four
quark operators because of the absence of $Z$-boson in the effective theory.
Therefore, the Wilson coefficient $C_{10}$ does not renormalize under QCD
corrections and is independent of the energy scale $\mu .$ Additionally the
above quark level decay amplitude can get contributions from the matrix
element of four quark operators, $\sum_{i=1}^{6}\left\langle
\ell^{+}\ell^{-}s\left\vert O_{i}\right\vert b\right\rangle ,$ which are usually
absorbed into the effective Wilson coefficient $C_{9}^{eff}(\mu )$ and can
be written as \cite{25, 26, 30, 31}
\begin{equation*}
\widetilde{C}_{9}^{eff}(\mu )=\widetilde{C}_{9}(\mu )+Y_{SD}(z,s^{\prime })+Y_{LD}(z,s^{\prime }).
\end{equation*}%
where $z=m_{c}/m_{b}$ and $s^{\prime }=s/m_{b}^{2}$. $Y_{SD}(z,s^{\prime
})$ describes the short distance contributions and the long distance contribution is $%
Y_{LD}(z,s^{\prime })$ . The manifest expressions of these contributions are given as:
\begin{eqnarray}
Y_{SD}(z,s^{\prime }) &=&h(z,s^{\prime })\left[3C_{1}(\mu )+C_{2}(\mu
)+3C_{3}(\mu )+C_{4}(\mu )+3C_{5}(\mu )+C_{6}(\mu )\right]  \notag \\
&&-\frac{1}{2}h(1,s^{\prime })\left[4C_{3}(\mu )+4C_{4}(\mu )+3C_{5}(\mu
)+C_{6}(\mu )\right]  \notag \\
&&-\frac{1}{2}h(0,s^{\prime })\left[C_{3}(\mu )+3C_{4}(\mu )\right]+{\frac{2}{9}}\left[3C_{3}(\mu )+C_{4}(\mu )+3C_{5}(\mu )+C_{6}(\mu )\right],
\end{eqnarray}%
\begin{eqnarray}
Y_{LD}(z,s^{\prime }) &=&\frac{3}{\alpha _{em}^{2}}(3C_{1}(\mu )+C_{2}(\mu
)+3C_{3}(\mu )+C_{4}(\mu )+3C_{5}(\mu )+C_{6}(\mu ))  \notag \\
&&\times\sum_{j=\psi ,\psi ^{\prime }}\omega _{j}(s)k_{j}\frac{\pi \Gamma
(j\rightarrow l^{+}l^{-})M_{j}}{s-M_{j}^{2}+iM_{j}\Gamma _{j}^{tot}},
\label{LD}
\end{eqnarray}%
with
\begin{eqnarray}
h(z,s^{\prime }) &=&-{\frac{8}{9}}\mathrm{ln}z+{\frac{8}{27}}+{\frac{4}{9}}x-%
{\frac{2}{9}}(2+x)|1-x|^{1/2}\left\{
\begin{array}{l}
\ln \left| \frac{\sqrt{1-x}+1}{\sqrt{1-x}-1}\right| -i\pi \quad \mathrm{for}{%
{\ }x\equiv 4z^{2}/s^{\prime }<1} \\
2\arctan \frac{1}{\sqrt{x-1}}\qquad \mathrm{for}{{\ }x\equiv
4z^{2}/s^{\prime }>1}%
\end{array}%
\right. ,  \notag \\
h(0,s^{\prime }) &=&{\frac{8}{27}}-{\frac{8}{9}}\mathrm{ln}{\frac{m_{b}}{\mu
}}-{\frac{4}{9}}\mathrm{ln}s^{\prime }+{\frac{4}{9}}i\pi \,\,.
\end{eqnarray}%
Here $M_{j}(\Gamma _{j}^{tot})$ are the masses (widths) of the intermediate
resonant states and $\Gamma (j\rightarrow l^{+}l^{-})$ denote the partial
decay width for the transition of vector charmonium state to massless lepton
pair, which can be expressed in terms of the decay constant of charmonium
through the relation \cite{32}
\begin{equation*}
\Gamma (j\rightarrow \ell^{+}\ell^{-})=\pi \alpha _{em}^{2}{\frac{16}{27}}{\frac{%
f_{j}^{2}}{M_{j}}}.
\end{equation*}%
The phenomenological parameter $k_{j}$ in Eq. (\ref{LD}) is to account for
inadequacies of the factorization approximation, and it can be determined
from
\begin{equation*}
{\cal BR}(B\rightarrow K_{1} J/\psi \rightarrow  K_{1}\ell^{+}\ell^{-})={\cal BR}(B\rightarrow K_{1} J/\psi )\cdot {\cal BR}(J/\psi\rightarrow \ell^{+}\ell^{-}).
\end{equation*}%
The function $\omega _{j}(s)$ introduced in Eq. (\ref{LD}) is to
compensate the naive treatment of long distance contributions due to the
charm quark loop in the spirit of quark-hadron duality, which can
overestimate the genuine effect of the charm quark at small $s$
remarkably \footnote{%
For extensive discussions on long-distance and short-distance
contributions from the charm loop, one can refer to the references \cite{32, b to s 2, b to s 3,charm loop 1, charm loop 2,charm
loop 3,yuming}.}. The quantity $\omega _{j}(s)$ can be normalized to $\omega
_{j}(M_{\psi _{j}}^{2})=1$, but its exact form is unknown at present. Since
the dominant contribution of the resonances is in the vicinity of the
intermediate $\psi _{i}$ masses, we will simply use $\omega _{j}(s)=1$
in our numerical calculations. In addition,  for the resonances $J/\psi$ and $\psi^{\prime}$
are taken to be $\kappa = 1.65$ and $\kappa = 2.36$, respectively \cite{IJA}.

Moreover, the non factorizable effects from the charm quark loop brings further
corrections to the radiative transition $b\rightarrow s\gamma ,$ and these
can be absorbed into the effective Wilson coefficients $C_{7}^{eff}$ which
then takes the form \cite{yuming,32, 33, 34, 35, 36}
\begin{equation*}
C_{7}^{eff}(\mu )=C_{7}(\mu )+C_{b\rightarrow s\gamma }(\mu )
\end{equation*}%
with
\begin{eqnarray}
C_{b\rightarrow s\gamma }(\mu ) &=&i\alpha _{s}\left[ \frac{2}{9}\eta
^{14/23}(G_{1}(x_{t})-0.1687)-0.03C_{2}(\mu )\right]  \label{8} \\
G_{1}(x_{t}) &=&\frac{x_{t}\left( x_{t}^{2}-5x_{t}-2\right) }{8\left(
x_{t}-1\right) ^{3}}+\frac{3x_{t}^{2}\ln ^{2}x_{t}}{4\left( x_{t}-1\right)
^{4}}  \label{9}
\end{eqnarray}%
where $\eta =\alpha _{s}(m_{W})/\alpha _{s}(\mu ),$ \ $%
x_{t}=m_{t}^{2}/m_{W}^{2}$ and $C_{b\rightarrow s\gamma }$ is the absorptive
part for the $b\rightarrow sc\bar{c}\rightarrow s\gamma $ rescattering.

\subsection{Form Factors and Mixing of $K_{1}(1270)-K_{1}(1400)$}\label{mix}

The exclusive $B\rightarrow K_{1}(1270,1400)\ell^{+}\ell^{-}$ decays involve the
hadronic matrix elements of quark operators given in Eq. (\ref{5a}). The different matrix elements
can be parameterized in terms of the form factors as:
\begin{align}
\left\langle K_{1}(k,\varepsilon )\left\vert V_{\mu }\right\vert
B(p)\right\rangle  =&\varepsilon _{\mu }^{\ast }\left(
M_{B}+M_{K_{1}}\right) V_{1}(s) -(p+k)_{\mu }\left( \varepsilon ^{\ast }\cdot q\right) \frac{V_{2}(s)}{%
M_{B}+M_{K_{1}}}  \notag \\
&-q_{\mu }\left( \varepsilon \cdot q\right)
\frac{2M_{K_{1}}}{s}\left[ V_{3}(s)-V_{0}(s)\right] \label{tf6} \\
\left\langle K_{1}(k,\varepsilon )\left\vert A_{\mu }\right\vert
B(p)\right\rangle  =&\frac{2i\epsilon _{\mu \nu \alpha \beta }}{%
M_{B}+M_{K_{1}}}\varepsilon ^{\ast \nu }p^{\alpha }k^{\beta }A(s)
\label{tf7}\\
\left\langle K_{1}(k,\varepsilon )\left\vert S\right\vert B_{c}(p)\right\rangle =&\mp \frac{2M_{K_{1}}}{m_{b}+m_{s}}\left(\varepsilon ^{\ast}\cdot p\right) V_{0}(s)\label{12}
\end{align}
where $V_{\mu }=\bar{s}\gamma _{\mu }b$, $A_{\mu }=\bar{s}\gamma_{\mu}\gamma _{5}b$ and $S=\bar{s}(1\pm\gamma_{5})b$ are the vector, axial vector and (pseudo)scalar currents, involved in the transition matrix, respectively. Also $p(k)$ are the momenta of the $B(K_{1})$ mesons, $q=p-k$ is the momentum transfer and $\varepsilon
_{\mu }$ correspond to the polarization of the final state axial vector
$K_{1}$ meson. In Eq. (\ref{tf6}), we have
\begin{equation}
V_{3}(s)=\frac{M_{B}+M_{K_{1}}}{2M_{K_{1}}}V_{1}(s)-\frac{M_{B}-M_{K_{1}}}{%
2M_{K_{1}}}V_{2}(s),  \label{tf8}
\end{equation}%
with
\begin{equation*}
V_{3}(0)=V_{0}(0).
\end{equation*}%
In addition, there is also a contribution from the
Penguin form factors which can be expressed as
\begin{align}
\left\langle K_{1}(k,\varepsilon )\left\vert \bar{s}i\sigma _{\mu
\nu }q^{\nu }b\right\vert B(p)\right\rangle  =&\left[ \left(
M_{B}^{2}-M_{K_{1}}^{2}\right) \varepsilon _{\mu }-(\varepsilon
\cdot
q)(p+k)_{\mu }\right] F_{2}(s)  \notag \\
&+(\varepsilon ^{\ast }\cdot q)\left[ q_{\mu }-\frac{s}{%
M_{B}^{2}-M_{K_{1}}^{2}}(p+k)_{\mu }\right] F_{3}(s)  \label{tf9}\\
\left\langle K_{1}(k,\varepsilon )\left\vert \bar{s}i\sigma
_{\mu \nu }q^{\nu }\gamma _{5}b\right\vert
B(p)\right\rangle=&-i\epsilon _{\mu \nu \alpha \beta }\varepsilon
^{\ast \nu }p^{\alpha }k^{\beta }F_{1}(s), \label{tf10}
\end{align}
with $F_{1}(0)=2F_{2}(0).$

As the physical states $K_{1}(1270)$ and $K_{1}(1400)$ are the mixture of $K_{1A}$ and $K_{1B}$ states with mixing angle $\theta_{K}$, as defined in Eqs. (\ref{mix1}-\ref{mix2}),  therefore, we can write
\begin{eqnarray}
\left(\begin{array}{c}\langle K_{1}(1270)\vert
\bar{s}\gamma_{\mu}(1-\gamma_{5})b\vert B\rangle\\
\langle K_{1}(1400)\vert \bar{s}\gamma_{\mu}(1-\gamma_{5})b\vert
B\rangle\end{array}\right)&=& M\left(\begin{array}{c}\langle
K_{1A}\vert
\bar{s}\gamma_{\mu}(1-\gamma_{5})b\vert B\rangle\\
\langle K_{1B}\vert \bar{s}\gamma_{\mu}(1-\gamma_{5})b\vert
B\rangle\end{array}\right),\label{m1}\\
\left(\begin{array}{c}\langle K_{1}(1270)\vert
\bar{s}\sigma_{\mu\nu}q^{\mu}(1+\gamma_{5})b\vert B\rangle\\
\langle K_{1}(1400)\vert
\bar{s}\sigma_{\mu\nu}q^{\mu}(1+\gamma_{5})b\vert
B\rangle\end{array}\right)&=& M\left(\begin{array}{c}\langle
K_{1A}\vert
\bar{s}\sigma_{\mu\nu}q^{\mu}(1+\gamma_{5})b\vert B\rangle\\
\langle K_{1B}\vert
\bar{s}\sigma_{\mu\nu}q^{\mu}(1+\gamma_{5})b\vert
B\rangle\end{array}\right),\label{m2}
\end{eqnarray}
where the mixing matrix $M$ is
\begin{equation}
M=\left(\begin{array}{cc}\sin\theta_{K}&\cos\theta_{K}\\
\cos\theta_{K}&-\sin\theta_{K}\end{array}\right).\label{m3}
\end{equation}

With these definitions, the corresponding form factors $A^{K_{1}}$, $V_{0,1,2}^{K_{1}}$ and
$F_{0,1,2}^{K_{1}}$  in $B\to K_{1}$  can be
parameterized in terms of the following relations
\begin{eqnarray}\left(\begin{array}{c}\frac{A^{K_{1}(1270)}}{m_{B}+m_{K_{1}(1270)}}\\
\frac{A^{K_{1}(1400)}}{m_{B}+m_{K_{1}(1400)}}\end{array}\right) &=&
M\left(\begin{array}{c}\frac{A^{K_{1A}}}{m_{B}+m_{K_{1A}}}\\
\frac{A^{K_{1B}}}{m_{B}+m_{K_{1B}}}\end{array}\right),\label{m4}\\
\left(\begin{array}{c}(m_{B}+m_{K_{1}(1270)})V_{1}^{K_{1}(1270)}\\
(m_{B}+m_{K_{1}(1400)})V_{1}^{K_{1}(1400)}\end{array}\right)&=&
M\left(\begin{array}{c}(m_{B}+m_{K_{1A}})V_{1}^{K_{1A}}\\
(m_{B}+m_{K_{1B}})V_{1}^{K_{1B}}\end{array}\right),\label{m5}\\
\left(\begin{array}{c}\frac{V_{2}^{K_{1}(1270)}}{m_{B}+m_{K_{1}(1270)}}\\
\frac{V_{2}^{K_{1}(1400)}}{m_{B}+m_{K_{1}(1400)}}\end{array}\right)
&=&
M\left(\begin{array}{c}\frac{V_{2}^{K_{1A}}}{m_{B}+m_{K_{1A}}}\\
\frac{V_{2}^{K_{1B}}}{m_{B}+m_{K_{1B}}}\end{array}\right),\label{m6}\\
\left(\begin{array}{c}m_{K_{1}(1270)}V_{0}^{K_{1}(1270)}\\
m_{K_{1}(1400)}V_{0}^{K_{1}(1400)}\end{array}\right) &=&
M\left(\begin{array}{c}m_{K_{1A}}V_{0}^{K_{1A}}\\
m_{K_{1B}}V_{0}^{K_{1B}}\end{array}\right),\label{m7}\\
\left(\begin{array}{c}F_{1}^{K_{1}(1270)}\\
F_{1}^{K_{1}(1400)}\end{array}\right) &=&
M\left(\begin{array}{c}F_{1}^{K_{1A}}\\
F_{1}^{K_{1B}}\end{array}\right),\label{m8}\\
\left(\begin{array}{c}(m^{2}_{B}-m^{2}_{K_{1}(1270)})F_{2}^{K_{1}(1270)}\\
(m^{2}_{B}+m^{2}_{K_{1}(1400)})F_{2}^{K_{1}(1400)}\end{array}\right)
&=&
M\left(\begin{array}{c}(m^{2}_{B}+m^{2}_{K_{1A}})F_{2}^{K_{1A}}\\
(m^{2}_{B}+m^{2}_{K_{1B}})F_{2}^{K_{1B}}\end{array}\right),\label{m9}\\
\left(\begin{array}{c}F_{3}^{K_{1}(1270)}\\
F_{3}^{K_{1}(1400)}\end{array}\right) &=&
M\left(\begin{array}{c}F_{3}^{K_{1A}}\\
F_{3}^{K_{1B}}\end{array}\right),\label{m10}
\end{eqnarray}
where we have supposed that $k^{\mu}_{K_{1}(1270),K_{1}(1400)}\simeq
k^{\mu}_{K_{1A},K_{1B}}$.
For the numerical analysis we have used the light-cone QCD sum rules form factors \cite{fmf}, summarized in Table
\ref{tabel1}, where the momentum dependence dipole parametrization is:
\begin{equation}
\mathcal{T}^{X}_{i}(s)=\frac{\mathcal{T}^{X}_{i}(0)}{1-a_{i}^{X}\left(s/m^{2}_{B}\right)+b_{i}^{X}\left(s/m^{2}_{B}\right)^{2}}\label{m11}.
\end{equation}
where $\mathcal{T}$ is $A$, $V$ or $F$ form factors and the subscript
$i$ can take a value 0, 1, 2 or 3 the superscript $X$  belongs to
$K_{1A}$ or $K_{1B}$ state.
\begin{table}[tbp]
\begin{tabular}{|p{.7in}p{.7in}p{.7in}p{.4in}||p{.7in}p{.7in}p{.7in}p{.4in}|}
\hline \hline
$\mathcal{T}^{X}_{i}(s)$&$\mathcal{T}(0)$&$a$&$b$&$\mathcal{T}^{X}_{i}(s)$&$\mathcal{T}(0)$&$a$&$b$\\
\hline
$V_{1}^{K_{1A}}$&$0.34$&$0.635$&$0.211$&$V_{1}^{K_{1B}}$&$-0.29$&$0.729$&$0.074$\\
$V_{2}^{K_{1A}}$&$0.41$&$1.51$&$1.18$&$V_{1}^{K_{1B}}$&$-0.17$&$0.919$&$0.855$\\
$V_{0}^{K_{1A}}$&$0.22$&$2.40$&$1.78$&$V_{0}^{K_{1B}}$&$-0.45$&$1.34$&$0.690$\\
$A^{K_{1A}}$&$0.45$&$1.60$&$0.974$&$A^{K_{1B}}$&$-0.37$&$1.72$&$0.912$\\
$F_{1}^{K_{1A}}$&$0.31$&$2.01$&$1.50$&$F_{1}^{K_{1B}}$&$-0.25$&$1.59$&$0.790$\\
$F_{2}^{K_{1A}}$&$0.31$&$0.629$&$0.387$&$F_{2}^{K_{1B}}$&$-0.25$&$0.378$&$-0.755$\\
$F_{3}^{K_{1A}}$&$0.28$&$1.36$&$0.720$&$F_{3}^{K_{1B}}$&$-0.11$&$1.61$&$10.2$\\
\hline\hline
\end{tabular}
\caption{$B\to K_{1A,1B}$ form factors \cite{fmf}, where $a$ and $b$
are the parameters of the form factors in dipole parametrization.}
\label{tabel1}
\end{table}

From Eq. (\ref{5a}), one can get the decay
amplitudes for $B\rightarrow K_{1}(1270)\ell^{+}\ell^{-} $  as
\begin{equation}
\mathcal{M}(B\rightarrow K_{1}\ell^{+}\ell^{-})=-\frac{%
G_{F}\alpha }{2\sqrt{2}\pi }V_{tb}V_{ts}^{\ast }
\left[ T_{V}^{\mu }\overline{%
\ell}\gamma _{\mu }\ell+T_{A}^{\mu }\overline{\ell}\gamma _{\mu }\gamma _{5}\ell+ T_{S}\left( \bar{\ell}\ell\right)\right]\label{35}
\end{equation}%
where the functions $T_{A}^{\mu }$ and $T_{V}^{\mu }$ can be written in terms of matrix elements as:
\begin{align}
T_{A}^{\mu } =&\widetilde{C}_{10}\left\langle K_{1}(k,\epsilon )\left\vert
\bar{s}\gamma ^{\mu }\left( 1-\gamma ^{5}\right)
b\right\vert B(p)\right\rangle   \\
T_{V}^{\mu } =&\widetilde{C}_{9}^{eff}\left\langle K_{1}(k,\epsilon )\left\vert
\bar{s}\gamma ^{\mu }\left( 1-\gamma ^{5}\right) b\right\vert %
B(p)\right\rangle -\widetilde{C}_{7}^{eff}\frac{2im_{b}}{s}\langle K_{1}(k,\epsilon
)\left\vert \bar{s}\sigma ^{\mu \nu }\left( 1+\gamma ^{5}\right)
q_{\nu }b\right\vert B(p)\rangle
\end{align}
which will take the form
\begin{align}
T_{V}^{\mu } =&f_{1}\epsilon
^{\mu \nu \rho \sigma }\varepsilon _{\nu}^{\ast }p_{\rho }k_{\sigma }-if_{2}%
\varepsilon ^{\ast\mu }-f_{3}(q\cdot\varepsilon)(p^{\mu }+k^{\mu })\label{TV}\\
T_{A}^{\mu } =& f_{4}\epsilon ^{\mu \nu \rho
\sigma }\varepsilon _{\mu}^{\ast}p_{\rho }k_{\sigma
}+if_{5}\varepsilon ^{\ast
\mu}-if_{6}(q\cdot\varepsilon)(p^{\mu }+k^{\mu })
+if_{7}(q\cdot\varepsilon)q^{\mu }
\label{VA}\\
 T_{S}=&2if_{8}(s)(\varepsilon ^{\ast }\cdot q)
\end{align}%

The auxiliary functions appearing in Eqs. (\ref{TV}) and (\ref{VA}) are defined as:
\begin{align}
f_{1}=&4(m_{b}+m_{s})\frac{\widetilde{C}_{7}^{eff}}{s}\bigg\{F_{1}^{K_{1A}}\sin\theta_{K}+F_{1}^{K_{1B}}\cos\theta_{K}\bigg\}\notag\\
&+2\widetilde{C}_{9}^{eff}\left\{\frac{A_{1}^{K_{1A}}\sin\theta_{K}}{m_{B}+m_{K_{1A}}}+\frac{A_{1}^{K_{1B}}\cos\theta_{K}}{m_{B}+m_{K_{1B}}}\right\}\label{ax-f1}\\
f_{2}=&2(m_{b}+m_{s})\frac{\widetilde{C}_{7}^{eff}}{s}\bigg\{(m_{B}^{2}-m^{2}_{K_{1A}})F_{2}^{K_{1A}}\sin{\theta_{K}}+(m_{B}^{2}-m^{2}_{K_{1B}})F_{2}^{K_{1B}}\cos{\theta_{K}}\bigg\}\notag\\
&+\widetilde{C}_{9}^{eff}\bigg\{(m_{B}+m_{K_{1A}})V_{1}^{K_{1A}}\sin{\theta_{K}}+(m_{B}+m_{K_{1B}})V_{1}^{K_{1B}}\cos{\theta_{K}}\bigg\}\label{aux-f2}\\
f_{3}=&2(m_{b}+m_{s})\frac{\widetilde{C}_{7}^{eff}}{s}\left\{\left(F_{2}^{K_{1A}}+\frac{sF_{3}^{K_{1A}}}{m_{B}^{2}-m^{2}_{K_{1A}}}\right)\sin{\theta_{K}}+\left(F_{2}^{K_{1B}}+\frac{sF_{3}^{K_{1B}}}{m_{B}^{2}-m^{2}_{K_{1B}}}\right)\cos{\theta_{K}}\right\}\notag\\
&+\widetilde{C}_{9}^{eff}\left(\frac{V_{2}^{K_{1A}}\sin{\theta_{K}}}{m_{B}+m_{K_{1A}}}+\frac{V_{2}^{K_{1B}}\cos{\theta_{K}}}{m_{B}+m_{K_{1B}}}\right)\\
f_{4}=&2\widetilde{C}_{10}^{eff}\left(\frac{A^{K_{1A}}\sin{\theta_{K}}}{m_{B}+m_{K_{1A}}}+\frac{A^{K_{1B}}\cos{\theta_{K}}}{m_{B}+m_{K_{1B}}}\right)\\
f_{5}=&\widetilde{C}_{10}\bigg\{(m_{B}+m_{K_{1A}})V_{1}^{K_{1A}}\sin{\theta_{K}}
+(m_{B}+m_{K_{1B}})V_{1}^{K_{1B}}\cos{\theta_{K}}\bigg\}\end{align}
\begin{align}
f_{6}=&\widetilde{C}_{10}\left(\frac{V_{2}^{K_{1A}}\sin{\theta_{K}}}{m_{B}+m_{K_{1A}}}+\frac{V_{2}^{K_{1B}}\cos{\theta_{K}}}{m_{B}+m_{K_{1B}}}\right)\\
f_{7}=&2\frac{\widetilde{C}_{10}}{s}\bigg\{m_{K_{1A}}\left(V_{3}^{K_{1A}}-V_{0}^{K_{1A}}\right)\sin{\theta_{K}}+m_{K_{1B}}\left(V_{3}^{K_{1B}}-V_{0}^{K_{1B}}\right)\cos{\theta_{K}}\bigg\}\notag\\
&+C_{Q_2}\bigg\{\frac{-m_{K_{1A}}V_{0}^{K_{1A}}\sin{\theta_{K}}+m_{K_{1B}}V_{0}^{K_{1B}}\cos{\theta_{K}}}{m_{\ell}(m_{b}+m_{s})}\bigg\}\\  \label{ax-f7}\\
f_{8}=&-C_{Q_1}\bigg\{\frac{-m_{K_{1A}}V_{0}^{K_{1A}}\sin{\theta_{K}}+m_{K_{1B}}V_{0}^{K_{1B}}\cos{\theta_{K}}}{m_{\ell}(m_{b}+m_{s})}\bigg\}\label{ax-f8}
\end{align}

\section{Physical Observables for $B\rightarrow K_{1} \ell^{+}\ell^{-}$}\label{obs}

In this section we will present the calculations of the physical observables
such as the branching ratios ${\cal BR}$, the forward-backward asymmetries ${\cal A}_{FB}$ and the lepton polarization asymmetries for the decays $B\rightarrow K_{1} \ell^{+}\ell^{-}$.
\subsection{Branching Ratio}

The double differential decay rate for $B\rightarrow K_{1}\ell^{+}\ell^{-}$
can be written as \cite{HY}%
\begin{equation}
\frac{d\Gamma (B\rightarrow K_{1}\ell^{+}\ell^{-})}{ds}=\frac{1}{\left( 2\pi \right) ^{3}}\frac{1}{32M_{B}^{3}}
\int_{-u(s)}^{+u(s)}du\left\vert \mathcal{M}\right\vert ^{2}
\label{62a}
\end{equation}%
where
\begin{eqnarray}
s &=&(p_{l^{+}}+p_{l^{-}})^{2} \\
u &=&\left( p-p_{l^{-}}\right) ^{2}-\left( p-p_{l^{+}}\right) ^{2}
\end{eqnarray}%
Now the limits on $s$ and $u$ are
\begin{eqnarray}
4m^{2} &\leq &s\leq (M_{B}-M_{K_{1}})^{2}  \label{62d} \\
-u(s) &\leq &u\leq u(s)  \label{62e}
\end{eqnarray}%
with%
\begin{equation}
u(s)=\sqrt{\lambda \left( 1-\frac{4m^{2}}{s}\right) }
\label{62f}
\end{equation}%
and%
\begin{equation*}
\lambda \equiv \lambda (M_{B}^{2},M_{K_{1}}^{2},s)=M_{B}^{4}+M_{K_{1}}^{4}+q^{4}-2M_{B}^{2}M_{K_{1}}^{2}-2M_{K_{1}}^{2}s-2sM_{B}^{2}
\end{equation*}%
Here $m$ corresponds to the mass of the lepton which for our case are
the $\mu$ and $\tau$. The total decay rate for the decay $B\rightarrow K_{1}\ell^{+}\ell^{-}$ can be expressed as
\begin{eqnarray}
\frac{d\Gamma}{ds} &=&\frac{G_{F}^{2}\left\vert
V_{tb}V_{ts}^{\ast }\right\vert ^{2}\alpha ^{2}}{2^{11}\pi
^{5}3M_{B}^{3}M_{K_{1}}^{2}s}u(s)\times \mathcal{A}\left(
s\right)  \label{62i}
\end{eqnarray}%
The function $u(s)$ is defined Eq. (\ref{62f}) and $\mathcal{M}(s)$ can be parametrized as
\begin{align}
\mathcal{A}(s) &=8M_{K_{1}}^{2}s\lambda
\bigg\{(2m^{2}+s)\left\vert f_{1}(s)\right\vert ^{2}
-(4m^{2}-s)\left\vert f_{4}(s)\right\vert ^{2}\bigg\}  +4M_{K_{1}}^{2}s\bigg\{(2m^{2}+s)\notag \\
&\times\left( 3\left\vert f_{2}(s)\right\vert ^{2}-\lambda \left\vert f_{3}(s)\right\vert
^{2}\right) -(4m^{2}-s)\left( 3\left\vert f_{5}(s)\right\vert ^{2}-\lambda
\left\vert f_{6}(s)\right\vert ^{2}\right) \bigg\}\notag \\
&+\lambda (2m^{2}+s)\left\vert
f_{2}(s)+\left(M_{B}^{2}-M_{K_{1}}^{2}-s\right)f_{3}(s)\right\vert ^{2}+ 24m^{2}M_{K_{1}}^{2}\lambda\left\vert f_{7}(s)\right\vert^{2} \notag \\
&-(4m^{2}-s)\left\vert f_{5}(s)+\left(M_{B}^{2}-M_{K_{1}}^{2}-s\right)f_{6}(s)\right\vert ^{2}  +\left(s-4 m^2\right) \lambda\left\vert f_{8}(s)\right\vert ^{2}  \notag \\
& -12 m^{2}s\left[\Re(f_{5}f_{7}^{\ast })-\Re(f_{6}f_{7}^{\ast })\right]\label{63b}\\
\end{align}

It is also very useful to define the ratio of the branching fractions $(mathcal{R}_{\ell})$ as:
\begin{equation}
\mathcal{R}_{\ell}=\frac{\mathcal{BR}(B\rightarrow K_{1}(1400)\ell^{+}\ell^{-})}{\mathcal{BR}(B\rightarrow K_{1}(1270)\ell^{+}\ell^{-})}
\end{equation}
where $\ell=\mu,\ \tau$.

\subsection{Forward-Backward Asymmetries}

In this section we investigate the forward-backward asymmetry ($\mathcal{A}_{FB}$)
of leptons. In the context of THD models, the $\mathcal{A}_{FB}$ can also play a crucial role in $B\rightarrow K_{1}\ell^{+}\ell^{-}$ transitions . The differential $\mathcal{A}_{FB}$ of final state lepton for the said decays can be
written as
\begin{equation}
{\frac{d\mathcal{A}_{FB}(s)}{ds}}=\int_{0}^{1}\frac{d^{2}\Gamma }{dsd\cos \theta }%
d\cos \theta -\int_{-1}^{0}\frac{d^{2}\Gamma }{dsd\cos \theta
}d\cos \theta  \label{FBformula}
\end{equation}%
From experimental point of view the normalized forward-backward
asymmetry is more useful, i.e., 
\begin{equation*}
\mathcal{A}_{FB}=\frac{\int_{0}^{1}\frac{d^{2}\Gamma }{dsd\cos \theta }%
d\cos \theta -\int_{-1}^{0}\frac{d^{2}\Gamma }{dsd\cos \theta
}d\cos \theta  }{\int_{-1}^{1}\frac{d^{2}\Gamma }{dsd\cos \theta
} d\cos \theta }
\end{equation*}%
The normalized $\mathcal{A}_{FB}$ for $B\rightarrow K_{1}\ell^{+}\ell^{-}$ can be obtained from Eq. (\ref{62a}) as%
\begin{eqnarray}
\mathcal{A}_{FB} &=& \frac{1}{d\Gamma /ds}\frac{G_{F}^{2}\alpha ^{2}}{2^{11}\pi ^{5}m_{B}^{3}}\left\vert V_{tb}V_{ts}^{\ast }\right\vert^{2}su(s)\bigg\{4{Re}[f_{2}^{\ast}f_{4}+f_{1}^{\ast }f_{5}]+2\lambda \Re[f_{3}^{\ast} f_{8}]\notag\\
&&+4\Re [f_{2}f_{8}^{\ast}] \left(-M_{B_{c}}^2+M_{D_{s}^{\ast }}^2+s\right)\bigg\}
 \label{FBA}
\end{eqnarray}%
where $d\Gamma /ds$ is given in Eq. (\ref{62i}).

\subsection{Single Lepton Polarization Asymmetries}


In the rest frame of the lepton and anti-lepton, the unit vectors along
longitudinal, normal and transversal component of the $\ell^{-}$ can be defined
as \cite{Aliev}:
\begin{subequations}
\begin{eqnarray}
s_{L}^{-\mu } &=&(0,\vec{e}^{-}_{L})=\left( 0,\frac{\vec{p}_{-}}{\left| \vec{p}%
_{-}\right| }\right) , \label{p-vectorsa} \\
s_{N}^{-\mu } &=&(0,\vec{e}^{-}_{N})=\left( 0,\frac{\vec{k} \times
\vec{p}_{-}}{\left| \vec{k}\times \vec{p}_{-}\right| }\right) ,
\label{p-vectorsb} \\
s_{T}^{-\mu } &=&(0,\vec{e}^{-}_{T})=\left( 0,\vec{e}_{N}\times \vec{e}%
_{L}\right) ,  \label{p-vectorsc}
\end{eqnarray}
\end{subequations}
where $\vec{p}_{-}$ and $\vec{k}$ are the three-momenta of the
lepton $\ell^{-}$ and $K_{1}$ meson, respectively, in the center mass
(c.m.) frame of $\ell^{+}\ell^{-}$ system. Lorentz transformation is used to boost
the longitudinal component of the lepton polarization to the c.m. frame of the
lepton pair as
\begin{equation}
\left( s_{L}^{-\mu }\right) _{CM}=\left( \frac{|\vec{p}_{-}|}{m},\frac{%
E\vec{p}_{-}}{m\left| \vec{p}_{-}\right| }\right)
\label{bossted component}
\end{equation}
where $E$ and $m$ are the energy and mass of the lepton. The normal
and transverse components remain unchanged under the Lorentz boost. The
longitudinal ($P_{L}$), normal ($P_{N}$) and transverse ($P_{T}$)
polarizations of lepton can be defined as:
\begin{equation}
P_{i}^{(\mp )}(s)=\frac{\frac{d\Gamma }{ds}(\vec{\xi}^{\mp }=\vec{e}%
^{\mp })-\frac{d\Gamma }{ds}(\vec{\xi}^{\mp }=-\vec{e}^{\mp })}{\frac{%
d\Gamma }{ds}(\vec{\xi}^{\mp }=\vec{e}^{\mp })+\frac{d\Gamma }{ds}(%
\vec{\xi}^{\mp }=-\vec{e}^{\mp })}  \label{polarization-defination}
\end{equation}%
where $i=L,\;N,\;T$ and $\vec{\xi}^{\mp }$ is the spin direction along the
leptons $\ell^{\mp }$. The differential decay rate for polarized lepton $\ell^{\mp
}$ in $B\rightarrow K_{1}\ell^{+}\ell^{-}$ decay along any spin
direction $\vec{\xi}^{\mp }$ is related to the unpolarized decay rate (\ref{62i}) with the following relation
\begin{equation}
\frac{d\Gamma (\vec{\xi}^{\mp })}{ds}=\frac{1}{2}\left( \frac{d\Gamma }{%
ds}\right) \left[1+(P_{L}^{\mp }\vec{e}_{L}^{\mp }+P_{N}^{\mp }\vec{e}%
_{N}^{\mp }+P_{T}^{\mp }\vec{e}_{T}^{\mp })\cdot \vec{\xi}^{\mp }\right].
\label{polarized-decay}
\end{equation}%
The expressions of the longitudinal, normal and transverse lepton
polarizations can be written as
\begin{align}
P_{L}(s) \propto &\frac{4\lambda}{3M_{K_{1}}^{2}}\sqrt{\frac{s-4m_{\ell}^{2}}{s}}\times  \bigg\{2\Re(f_{2}f_{5}^{\ast})+\lambda\Re(f_{3}f_{6}^{\ast})+4\sqrt{s}\Re(f_{1}f_{4}^{\ast})\left(1+\frac{12sM_{K_{1}}^{2}}{\lambda}\right)\notag\\
& +\left(-M_{B}^{2}+M_{K_{1}}^{2}+s\right)\left[\Re(f_{3}f_{5}^{\ast})+\Re(f_{2}f_{6}^{\ast})\right]\notag\\
&+ \frac{3}{2}m_{\ell} \left[\Re(f_{5}f_{8}^{\ast})+\Re(f_{6}f_{8}^{\ast})\left(-M_{B}^{2}+M_{K_{1}}^{2}\right)-\Re(f_{7}f_{8}^{\ast})\right]\bigg\}\label{long-polarization}
\end{align}
\begin{align}
P_{N}(s) \propto & \frac{m_{\ell} \pi}{M_{K_{1}}^{2}}\sqrt{\frac{\lambda}{s}}\times \bigg\{-\lambda s\Re(f_{3}f_{7}^{\ast})+\lambda(M_{B}^{2}-M_{K_{1}}^{2})\Re(f_{3}f_{6}^{\ast})-\lambda\Re(f_{3}f_{5}^{\ast})\notag\\
&+\left(-M_{B}^{2}+M_{K_{1}}^{2}+s\right)\left[s\Re(f_{2}f_{7}^{\ast})+(M_{B}^{2}-M_{K_{1}}^{2})\Re(f_{2}f_{5}^{\ast})+(s-4m^2)\Re(f_{5}f_{8}^{\ast})\right]\notag\\
& - 8sM_{K_{1}}^{2}\Re(f_{1}f_{2}^{\ast}) +\sqrt{\lambda }(s-4m^2)\Re(f_{6}f_{8}^{\ast})\bigg\}   \label{norm-polarization}\\
P_{T}\left( s\right)  \propto & i\frac{m_{\ell} \pi \sqrt{ \left( s-\frac{4m_{l}^{2}}{s}\right) \lambda }}{M_{K_{1}}^{2}}\bigg\{M_{K_{1}}\left[4\Im(f_{2}f_{4}^{\ast})+4\Im(f_{1}f_{5}^{\ast})+3\Im(f_{5}f_{6}^{\ast})\right]-\lambda\Im(f_{6}f_{7}^{\ast})\notag\\
&+\left(-M_{B}^{2}+M_{K_{1}}^{2}+s\right)\bigg[\Im(f_{7}f_{5}^{\ast})+\Im(f_{2}f_{8}^{\ast})\bigg]-s\Im(f_{5}f_{6}^{\ast})\bigg\}
\label{Transverse-polarization}
\end{align}%
where the auxiliary functions $f_{1},f_{2},\cdots,f_{8}$ are defined in Eqs. (\ref{ax-f1})-(\ref{ax-f8}). Here we have dropped out the constant factors which are, however, understood.




\section{Numerical Results and Discussion}\label{num}

In order to perform the numerical analysis of the forward-backward asymmetry $(\mathcal{A}_{FB})$ and the lepton polarizations asymmetries $P_{L,N,T}$ for the $B\rightarrow
K_{1}(1270)\ell^{+}\ell^{-}$ decays, with $\ell=\mu,\tau$, we first give the
numerical values of input parameters and the SM Wilson coefficients which are used in our numerical
calculations in Tables \ref{input} and III, respectively. In principle, the above listed asymmetries can also be studied when we have $K_{1}(1400)$ meson instead of $K_{1}(1270)$ meson in the 
final state. It has already been pointed in literature \cite{aqeel} that the branching ratio of $B \to K_{1}(1400) \ell^{+}\ell^{-}$ is an order of magnitude smaller than its partner $B \to K_{1}(1270) \ell^{+}\ell^{-}$ decay, therefore, we will limit our study to the case when $K_1(1270)$ meson comes in the final state.
\begin{table}[ht]
\centering
\begin{tabular}{l}
\hline\hline
$m_{B}=5.28$ GeV, $m_{b}=4.28$ GeV, $m_{\mu}=0.105$ GeV,\\
$m_{\tau}=1.77$ GeV, $f_{B}=0.25$ GeV, $|V_{tb}V_{ts}^{\ast}|=45\times
10^{-3}$,\\ $\alpha^{-1}=137$, $G_{F}=1.17\times 10^{-5}$ GeV$^{-2}$,\\
$\tau_{B}=1.54\times 10^{-12}$ sec, $m_{K_{1}(1270)}=1.270$ GeV,\\
$m_{K_{1}(1400)}=1.403$ GeV, $\theta_{K}=-34^{\circ}$, \\ $m_{K_{1A}}=1.31$ GeV, $m_{K_{1B}}=1.34$ GeV \cite{KCyang}.\\
\hline\hline
\end{tabular}
\caption{Values of input parameters used in our numerical analysis \cite{pdg}.}\label{input}
\end{table}

\begin{table*}[ht]
\centering \caption{The Wilson coefficients $C_{i}^{\mu}$ at the
scale $\mu\sim m_{b}$ in the SM .}
\begin{tabular}{cccccccccc}
\hline\hline
$C_{1}$&$C_{2}$&$C_{3}$&$C_{4}$&$C_{5}$&$C_{6}$&$C_{7}$&$C_{9}$&$C_{10}$
\\ \hline
 \ \  1.107 \ \  & \ \  -0.248 \ \  & \ \  -0.011 \ \  & \ \  -0.026 \ \  & \ \  -0.007 \ \  & \ \  -0.031 \ \  & \ \  -0.313 \ \  & \ \  4.344 \ \  & \ \  -4.669 \ \  \\
\hline\hline
\end{tabular}
\label{wc table}
\end{table*}
Of course to perform the numerical analysis, another important ingredient is the form factors. The values of the form factors used in the upcoming analysis are
the ones calculated using the QCD sum rules and these are summarized in Table I.

Coming to the THDM, the free parameters in these models are the masses of charged Higgs boson $m_{H^{\pm}}$, the coefficients $\lambda_{tt}$, $\lambda_{bb}$ and the ratio of the vacuum expectation values of the two Higgs doubles, i.e. $tan\beta$. The coefficients $\lambda_{tt}$ and $\lambda_{bb}$ for the version I and II of the THDM are:
\begin{eqnarray}
\lambda_{tt} = \cot\beta,\hskip 1em  \lambda_{bb} = - \cot\beta, \hskip 1em \textit{for model I}, \notag \\
\lambda_{tt} = \cot\beta, \hskip 1em \lambda_{bb} =  +\tan\beta, \hskip 1em \textit{for model II}. \label{lambdas}
\end{eqnarray}
and for version III of THDM, these coefficients are complex, i.e.,
\begin{equation}
\lambda_{tt}\lambda_{bb} \equiv |\lambda_{tt}\lambda_{bb} | e^{i \delta},
\end{equation}
where $\delta$ is a single $CP$ phase of the vacuum in this version.

The constraints on the mass of charged Higgs boson and $\tan \beta$ are usually obtained by using the experimental results of the branching ratio of $b\to s \gamma$ and $B \to D \ell \nu_{\ell}$ decays as well as $B - \bar{B}$ and $K - \bar{K}$ mixing in the literature \cite{constraintsTHDM}. In addition the parameters 
$|\lambda_{tt}|$,  $\lambda_{bb}|$ and the phase $\delta$ are restricted by the experimental results of the electric dipole moments of neutron, $B - \bar{B}$ mixing, 
$\rho_0$, $R_b$ and $Br(b \to s \gamma)$ \cite{constraint1, constraint2, constraint3, constraint4}. The value of $\lambda_{tt}\lambda_{bb}$ is constrained to be 1 and the $\delta$ is restricted in the range $60^{\circ} - 90^{\circ}$ by using the experimental limits on electric dipole moment of neutron and $Br(b \to s \gamma)$, plus the constraint on $M_{H^{+}}$ from the LEP II. Using the constraints from the $B - \bar{B}$ mixing as well as from $R_b$, the analysis of various lepton polarization asymmetries in $B \to K_0^{\ast} \ell^{+}\ell^{-}$ has been done in the following parametric space in model III \cite{Falahati}:
\begin{eqnarray}
\textit{Case A}: \hskip 1em |\lambda_{tt}| = 0.03,\hskip 1em  |\lambda_{bb}| = 100, \hskip 1em \notag \\
\textit{Case B}: \hskip 1em |\lambda_{tt}| = 0.15,\hskip 1em  |\lambda_{bb}| = 50, \hskip 1em \label{lamdavalues} \\
\textit{Case C}: \hskip 1em |\lambda_{tt}| = 0.3,\hskip 1em  |\lambda_{bb}| = 30. \hskip 1em \notag
\end{eqnarray}
Where $\delta =\pi/2$ and the values of masses of Higgs particles are summarized in Table 4:
\begin{table}[ht]
\centering
\begin{tabular}{lcccc}
\hline\hline
Masses & $m_{A^0}$&  $m_{h^0}$& $m_{H^0}$& $m_{H^\pm}$ \\
\hline
Set I (GeV) &  $125$& $125$&  $160$& $200$\\
Set II (GeV) & $125$&  $125$& $160$& $160$\\
\hline\hline
\end{tabular}
\caption{Values of the masses of the Higgs particles.}\label{susy2}
\end{table}

It is an established fact that in THDM of type II the charged Higgs contribution to $B \to \tau \nu$ interferes necessarily destructive with the SM \cite{wu}. The enhancement of $Br(B \to \tau \nu)$ is possible if the absolute value of the contribution of the charged Higgs boson is two times the SM one, but then it is in conflict with the $B \to D \tau \nu$. Furthermore, this version of THDM can not explain the observed discrepancy of $2.2\sigma$ in $R(D)$ and $2.7\sigma$ in $R(D^*)$ compared to their SM value. In order to cure this situation, a detailed discussion on the model III has been done in Ref. \cite{Greub}. The purpose of present study is not to put the precise bounds on the parameters of versions of THDM but is to check the profile of different physical observables, e.g. the lepton forward-backward asymmetry as well as the lepton polarization asymmetries in $B \to K_{1} \ell^{+} \ell^{-}$ decays.  

It is important to mention here that as an exclusive decay, there are different source of uncertainties involved in the analysis of the above mentioned decay. The major source of uncertainties in the numerical analysis of $B \to K_1 \ell^+ \ell^-$ ($\ell=\mu,\tau$) decays originated from the $B \to K_1$ transition form factors summarized in Table 1. But it is also important to stress that these hadronic uncertainties have almost no influence on the various asymmetries including the forward-backward asymmetries and the lepton polarization asymmetries in $B \to K_1\ell^{+}\ell^{-}$ because of the cancellation among different polarization states and this make them a good tool to probe for physics beyond the SM.

\subsection{Analysis of Forward-Backward Asymmetry}
\begin{figure}[ht]
\begin{tabular}{cc}
\centering
\hspace{0.5cm}($\mathbf{a}$)&\hspace{1.2cm}($\mathbf{b}$)\\
\includegraphics[scale=0.6]{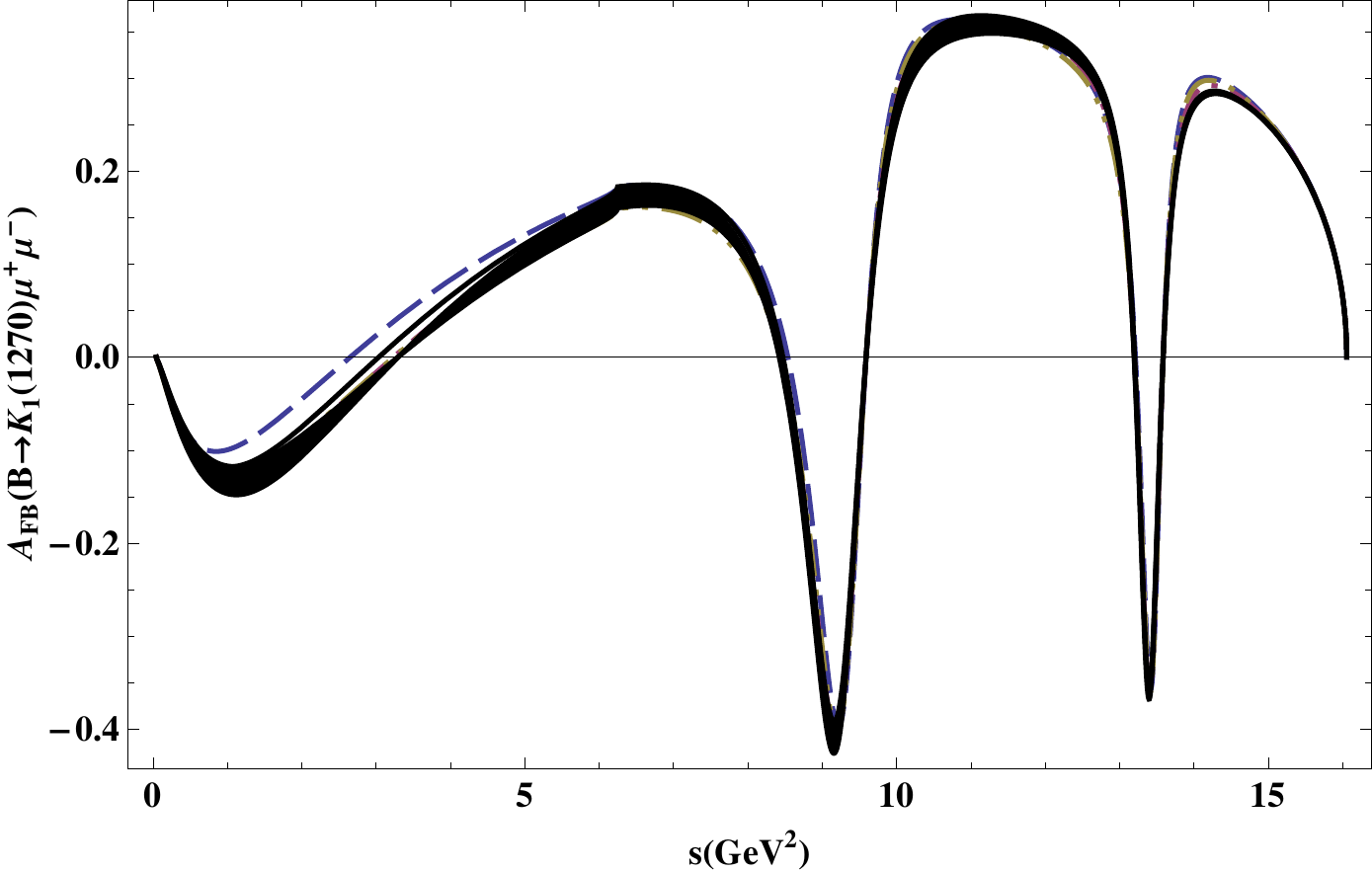}&\includegraphics[scale=0.6]{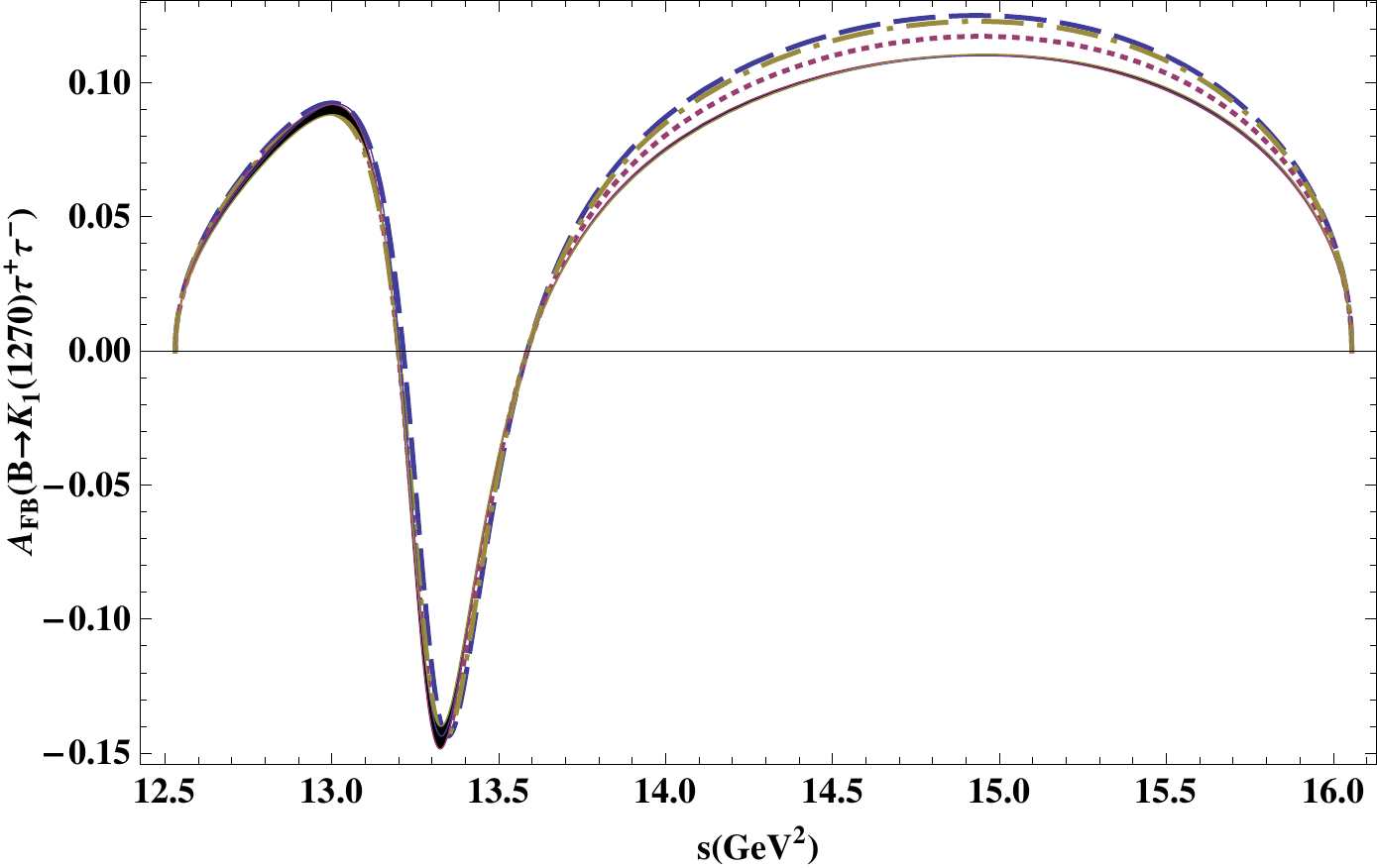}\\
\hspace{0.5cm}($\mathbf{c}$)&\hspace{1.2cm}($\mathbf{d}$)\\
\includegraphics[scale=0.6]{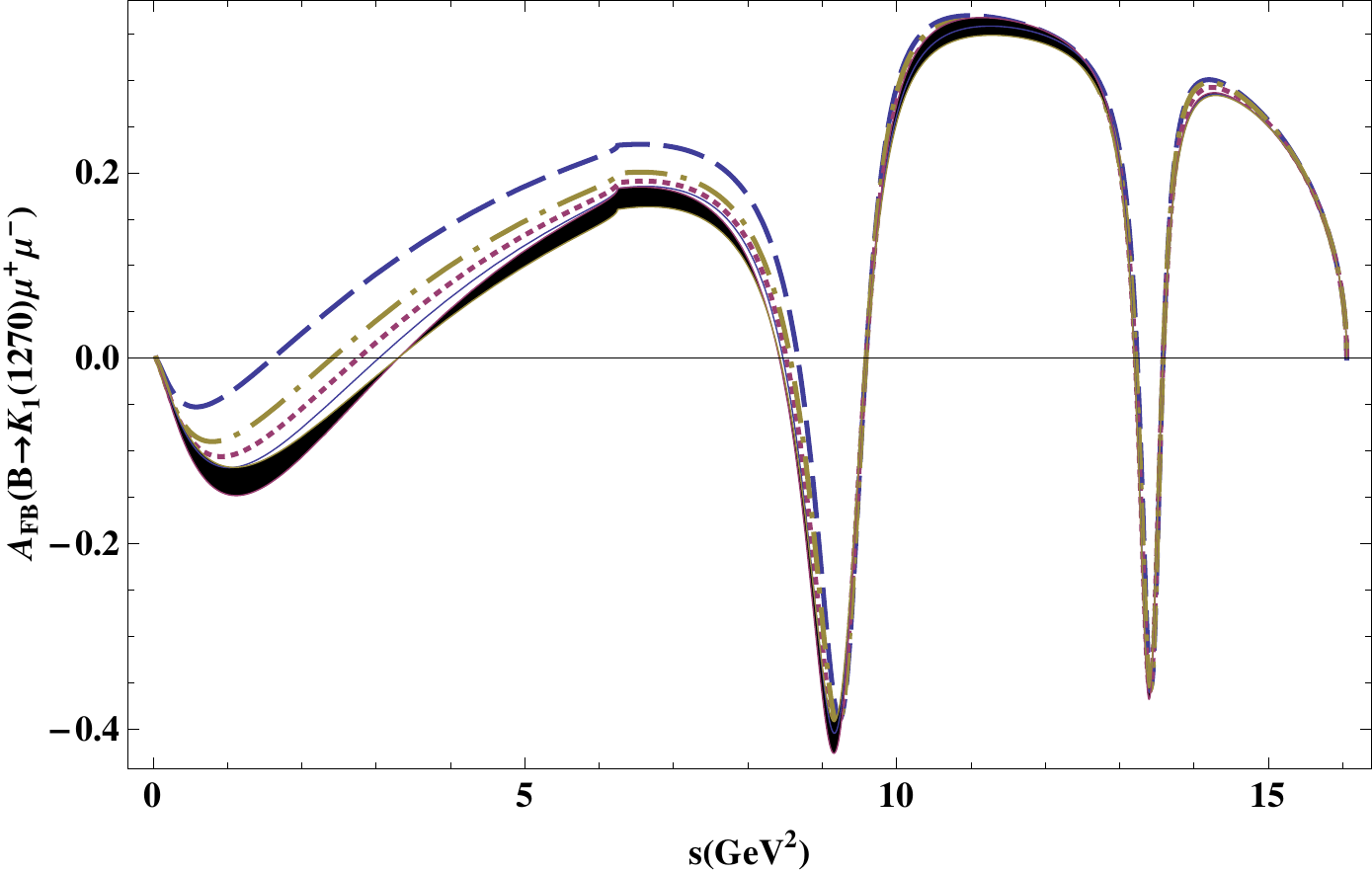}&\includegraphics[scale=0.6]{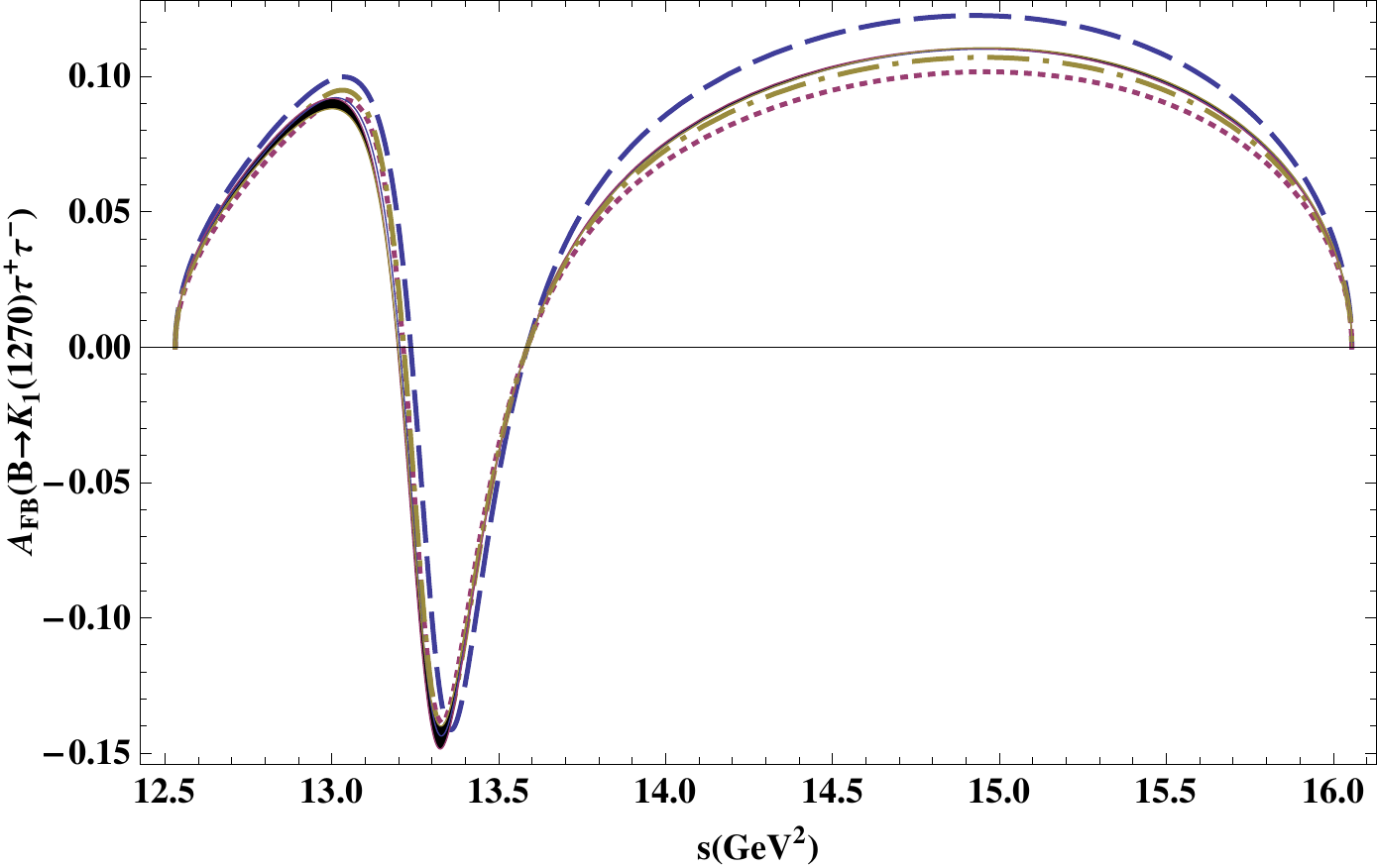}\end{tabular}
\caption{The dependence of forward-backward asymmetry of $B\to K_{1}\ell^{+}\ell^{-}$ on $s$ with long-distance contributions for (a) muons and (b) tauons in THDM1 and for (c) muons and (d) tauons in THDM2. In all the graphs solid band corresponds to the SM result along with uncertainties. Thea dashed, dotted and dashed-dot lines correspond to the case when mass of charged Higgs boson $m_{H^\pm} = 300 GeV, 400 GeV $ and $500 GeV$ respectively. The values of other parameters are $\lambda_{tt}=1$, $\lambda_{bb}=-1$ in model I and $\lambda_{tt}=1$, $\lambda_{bb}=1$ model II, whereas $m_{H^0} = 500 GeV$ and $m_{h^0} = 125 GeV$.} \label{brm1}
\end{figure}

To illustrate the impact of the parametric space of the THDM  on the forward-backward asymmetry $\mathcal{A}_{FB}$, we plot
$\frac{d(\mathcal{A}_{FB})}{ds}$ as a function of $s$ in Fig. \ref{brm1}. It is argued for the zero position of $\mathcal{A}_{FB}$ that the uncertainty in its position due to the hadronic form factors is negligible \cite{new2}. Therefore, the zero position of the $\mathcal{A}_{FB}$ can serve as a stringent test for the NP effects arising from the different versions of THDM. Figures \ref{brm1}(a) and \ref{brm1}(c) describe the ${\cal A}_{FB}$ for $B \to K_1 \mu^{+}\mu^{-}$ with long-distance contributions in the Wilson coefficients both for the THDM types I and II, respectively. Before, we discuss the attitude of different parameters in the forward-backward asymmetry it will be useful to give a closer look to Eqs. (\ref{C7mw}, \ref{C9mw}, \ref{C10mw}). In order to recover the SM phenomenology one has to put the parameter $y=0$. It can been seen that in THDM type I, because of the different sign of the $\lambda_{tt}$ and $\lambda_{bb}$ the second term in Eq. (\ref{C7mw}) gives constructive contributions where as the third term gives destructive contribution. Contrary to this the effects in Wilson coefficient $C_7$ in type II model are constructive and hence we expect large deviation form the SM value in this version compared to that of type I. In $B \to K_1 \mu^+ \mu^-$ decay, Figs. \ref{brm1}(a) and \ref{brm1}(c) depict this fact. Here, we can see that in case of type I the deviation of the zero position of the forward-backward asymmetry lies almost in the uncertainty band, since the only contribution of NP is coming in the Wilson coefficient $C_7$. However, in case of type II, the zero position as well as the magnitude of the $\mathcal{A}_{FB}$ shifted significantly from the SM value, especially when we have changed the values of the charged Higgs mass in this version.  Therefore, the precise measurement of the zero position of $\mathcal{A}_{FB}$ for the decay $B\to K_1\mu^{+}\mu^{-}$ will be a very good observable to yield any indirect imprints of NP due to the parameters of THDM and can serve as a good tool to distinguish among the different variants of it. In addition, the situation for $B \to K_1 \tau^+ \tau^-$ is shown in Figs. \ref{brm1}(b) and \ref{brm1}(d) where the shift in the value of $\mathcal{A}_{FB}$ is small compared to the case when we have $\mu$'s as final state leptons.  

\begin{figure}[ht]
\begin{tabular}{cc}
\centering
\hspace{0.5cm}($\mathbf{a}$)&\hspace{1.2cm}($\mathbf{b}$)\\
\includegraphics[scale=0.6]{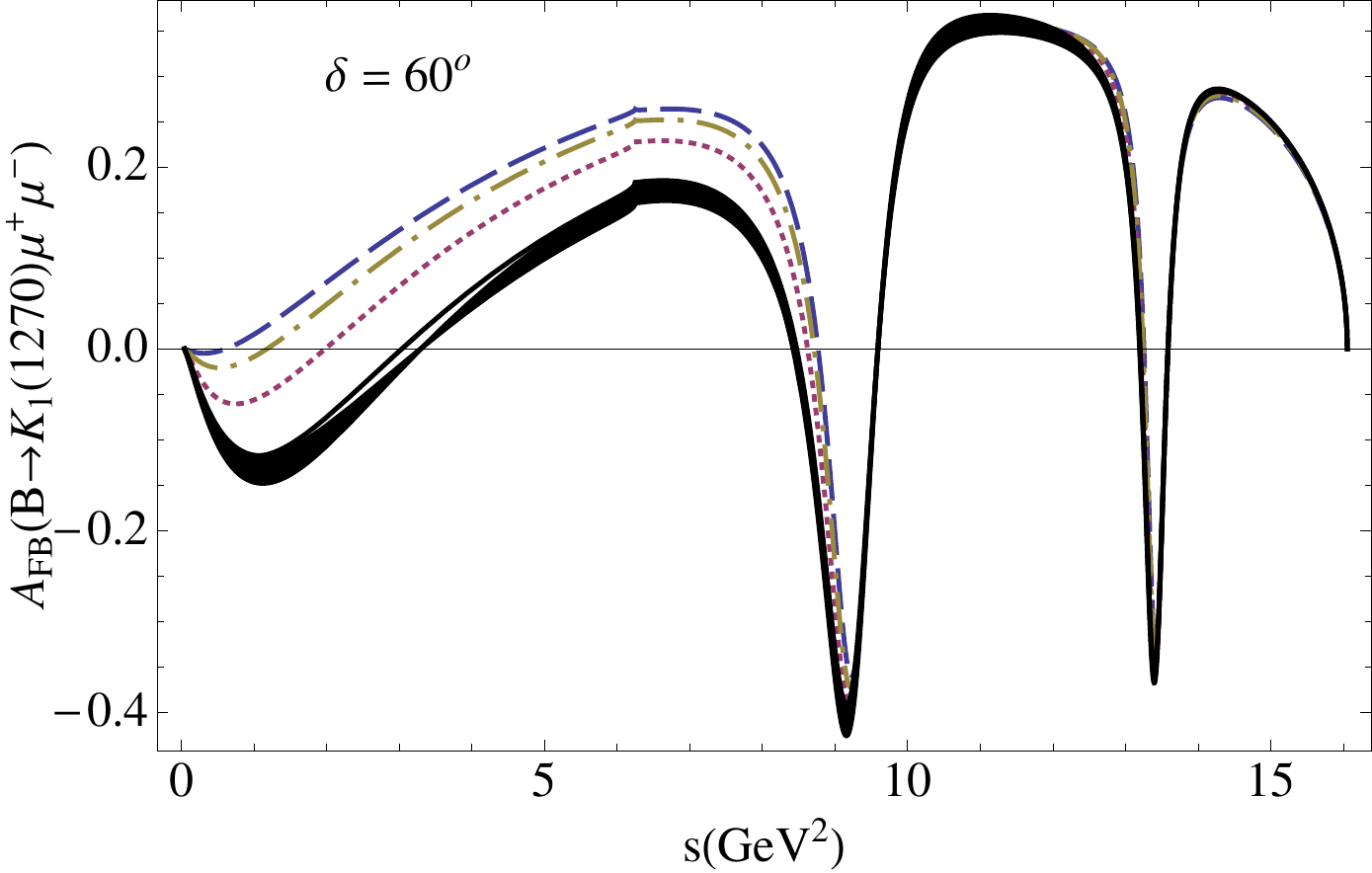}&\includegraphics[scale=0.6]{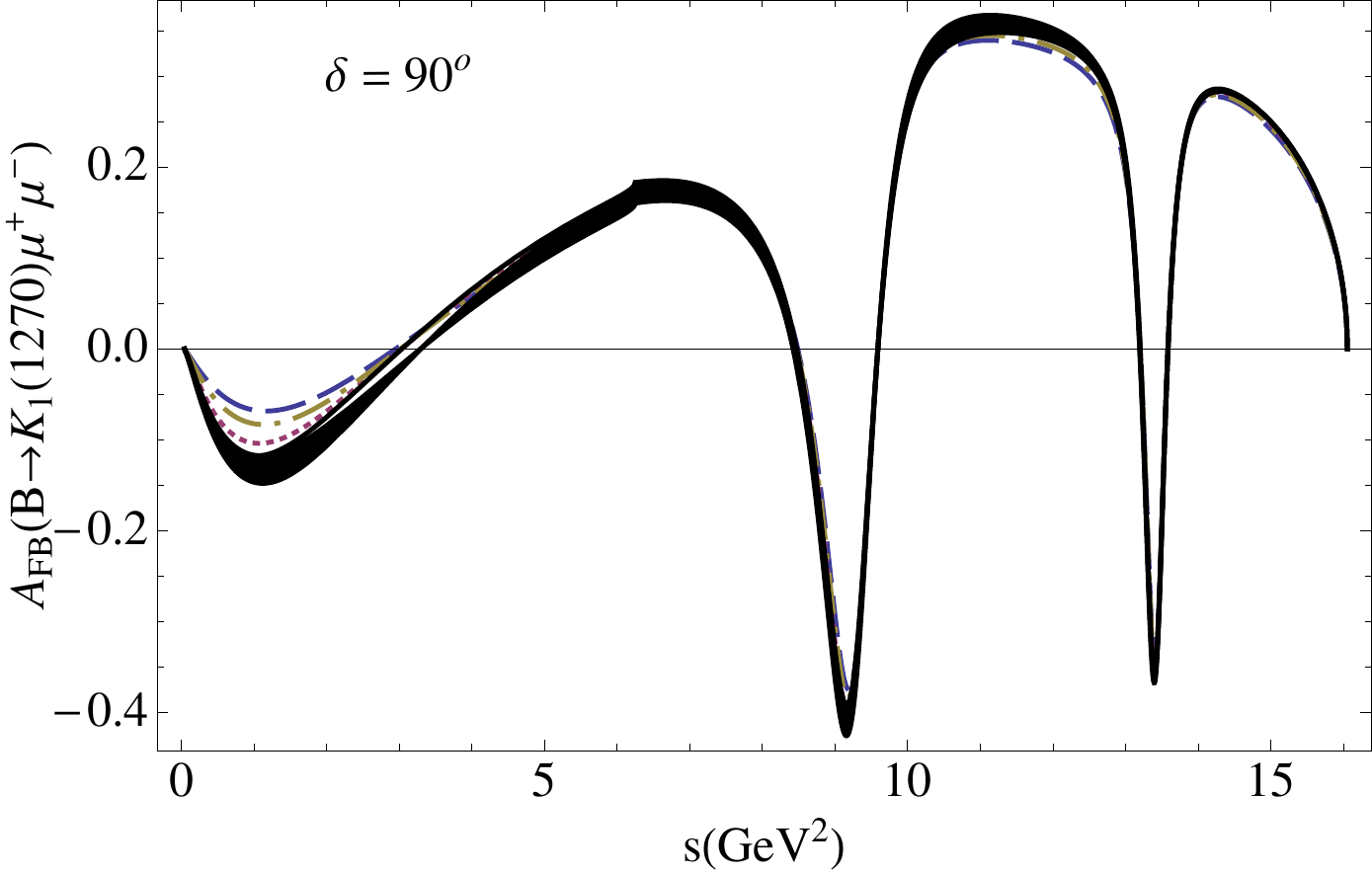}\\
\hspace{0.5cm}($\mathbf{c}$)&\hspace{1.2cm}($\mathbf{d}$)\\
\includegraphics[scale=0.6]{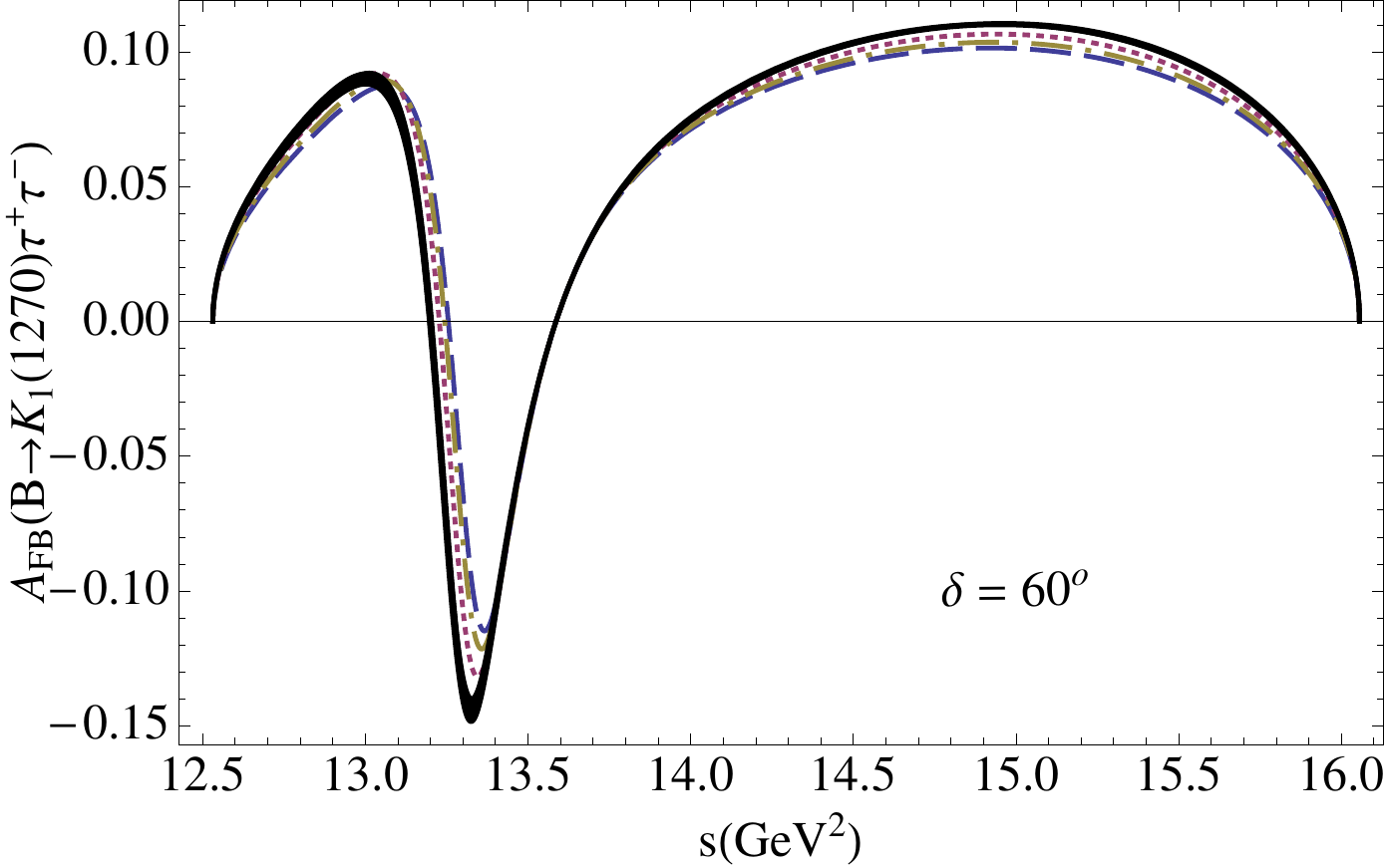}&\includegraphics[scale=0.6]{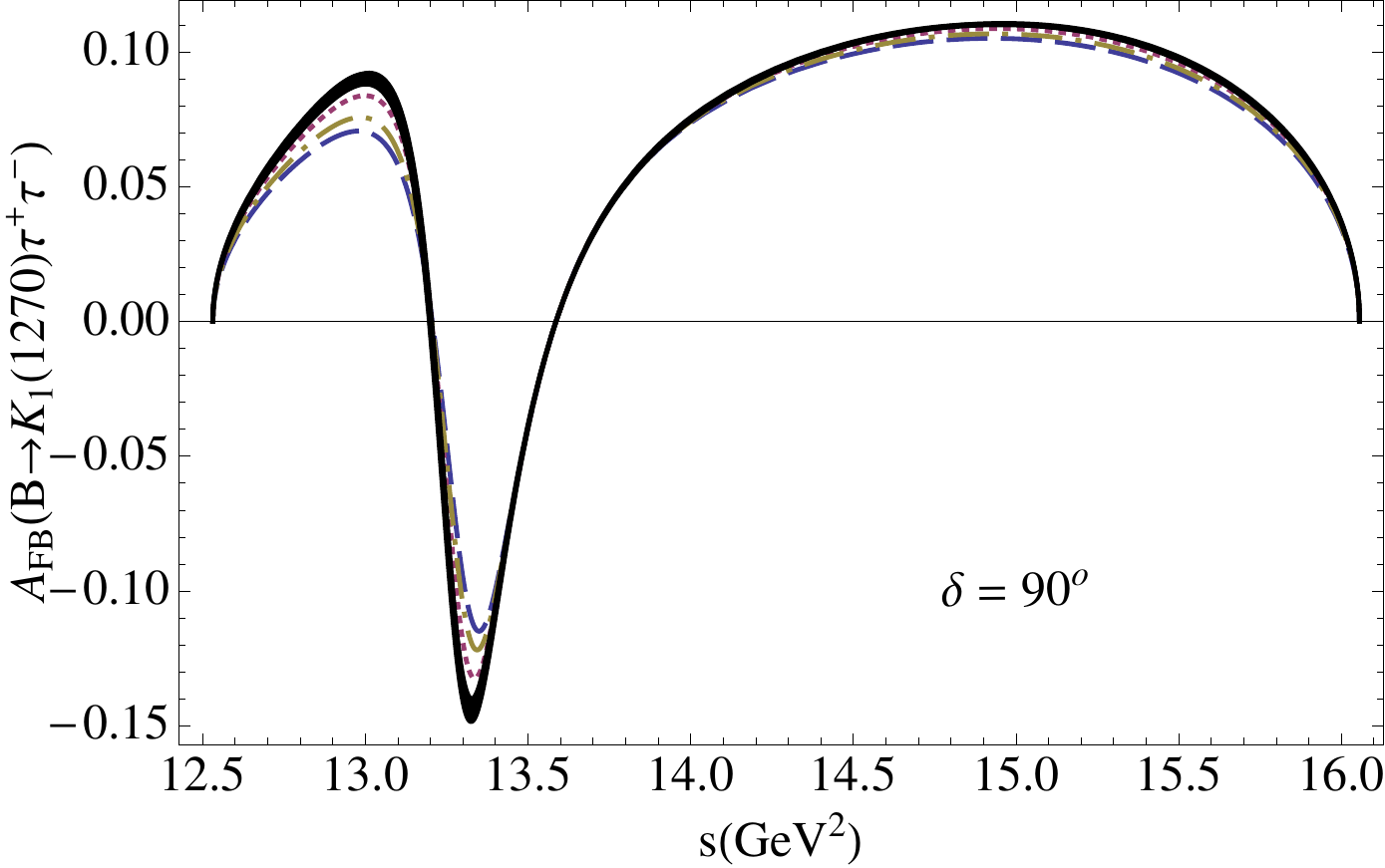}\end{tabular}
\caption{The dependence of forward-backward asymmetry of $B\to K_{1}\ell^{+}\ell^{-}$ on $s$ with long-distance contributions for muons (a) $\delta=60^\circ$ (b) $\delta=90^\circ$  and for tauons (c) $\delta=60^\circ$ (d) $\delta=90^\circ$ in THDM of type III. In all the graphs solid band corresponds to the SM uncertainties. The dashed, dotted and dashed-dot lines correspond Case A, Case B and Case C (c.f. Eq. (\ref{lamdavalues})), respectively.} \label{brm2}
\end{figure}

\begin{figure}[ht]
\centering
\includegraphics[scale=0.7]{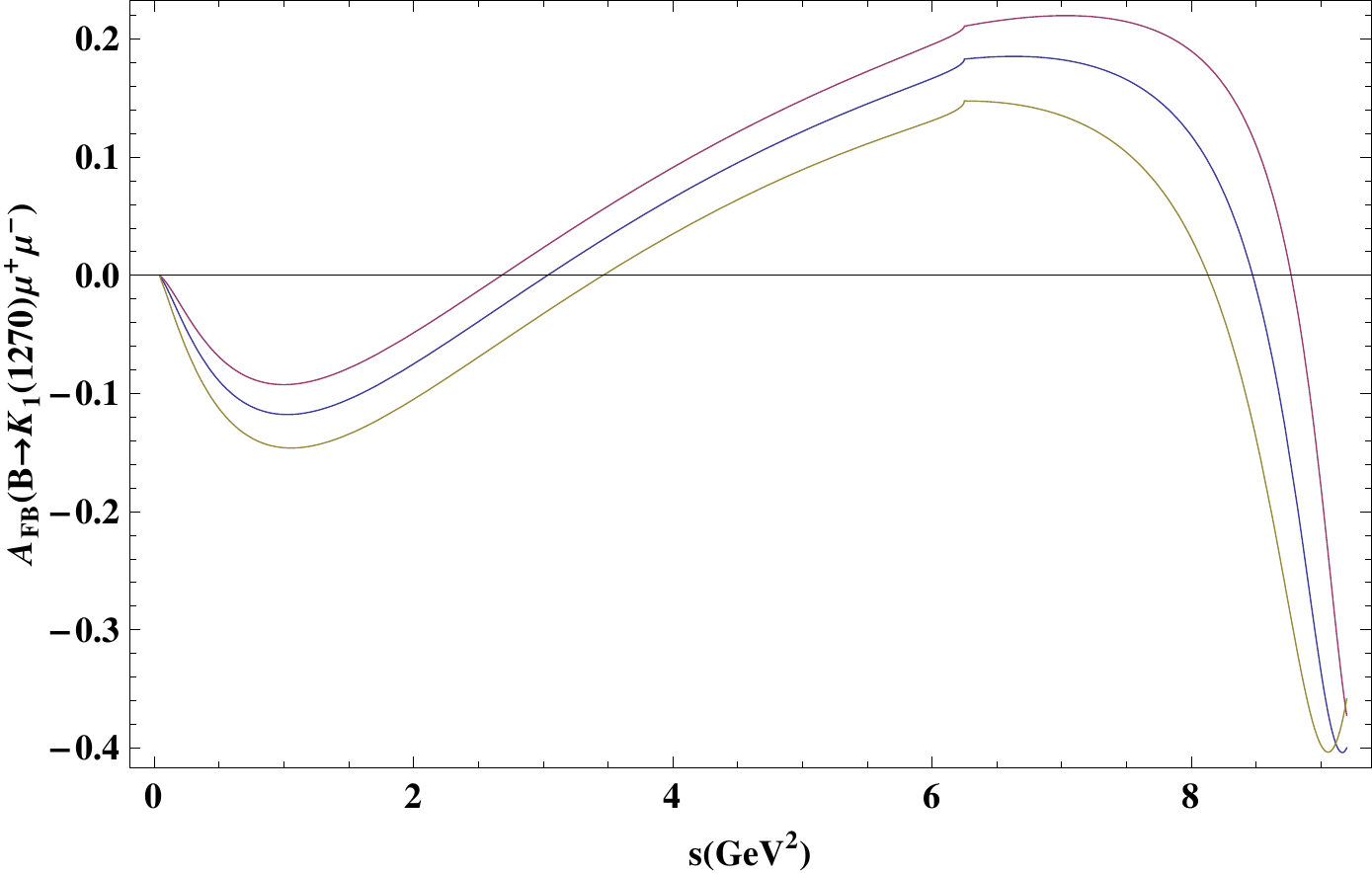}
\caption{The dependence forward-backward asymmetry on $s$ after including the charm-quark loop effects. The blue line corresponds to the case when charm-quark loop is ignored, where is purple and golden lines correspond to the case when $\delta_i$ in Eq. (\ref{cloop}) is taken to be 1 and $-1$, respectively. } \label{cloop-figure}
\end{figure}

In Fig. \ref{brm2}, the effects of different parameters corresponding to type III of THDM are shown in the FB-asymmetry in $B \to K_1 \ell^+ \ell^-$ $(\ell= \mu, \tau)$. It can be seen in Fig. \ref{brm2}(a) and \ref{brm2}(b) that in case of $\mu's$ as final state leptons the zero position of FB-asymmetry is sensitive to the phase angle $\delta$ and for $\delta=60^\circ$ this shift is maximum for maximum value of $\lambda_{bb}\lambda_{tt}$. This effect can easily be understood if we closely look at Eq. (\ref{FBA}), where it can be seen that $\mathcal{A}_{FB}$ is proportional to the real part of the combination of auxiliary functions $f_2$ and $f_8$. In Eq. (\ref{aux-f2}) the term proportional $C_7$ involves the new phase $\delta$ and for $\delta=90^\circ$ the NP contribution coming to $C_{7}$ is zero and hence the deviation form the SM is small compared to the case when $\delta=60^\circ$. Contrary to the $\mu$ case, the NP effects in $\mathcal{A}_{FB}$ for $B \to K_1\tau^+\tau^-$ are too faint for the whole range of phase $\delta$ and other parameters of the THDM type III.

\begin{figure}[ht]
\begin{tabular}{cc}
\centering
\hspace{0.5cm}($\mathbf{a}$)&\hspace{1.2cm}($\mathbf{b}$)\\
\includegraphics[scale=0.5]{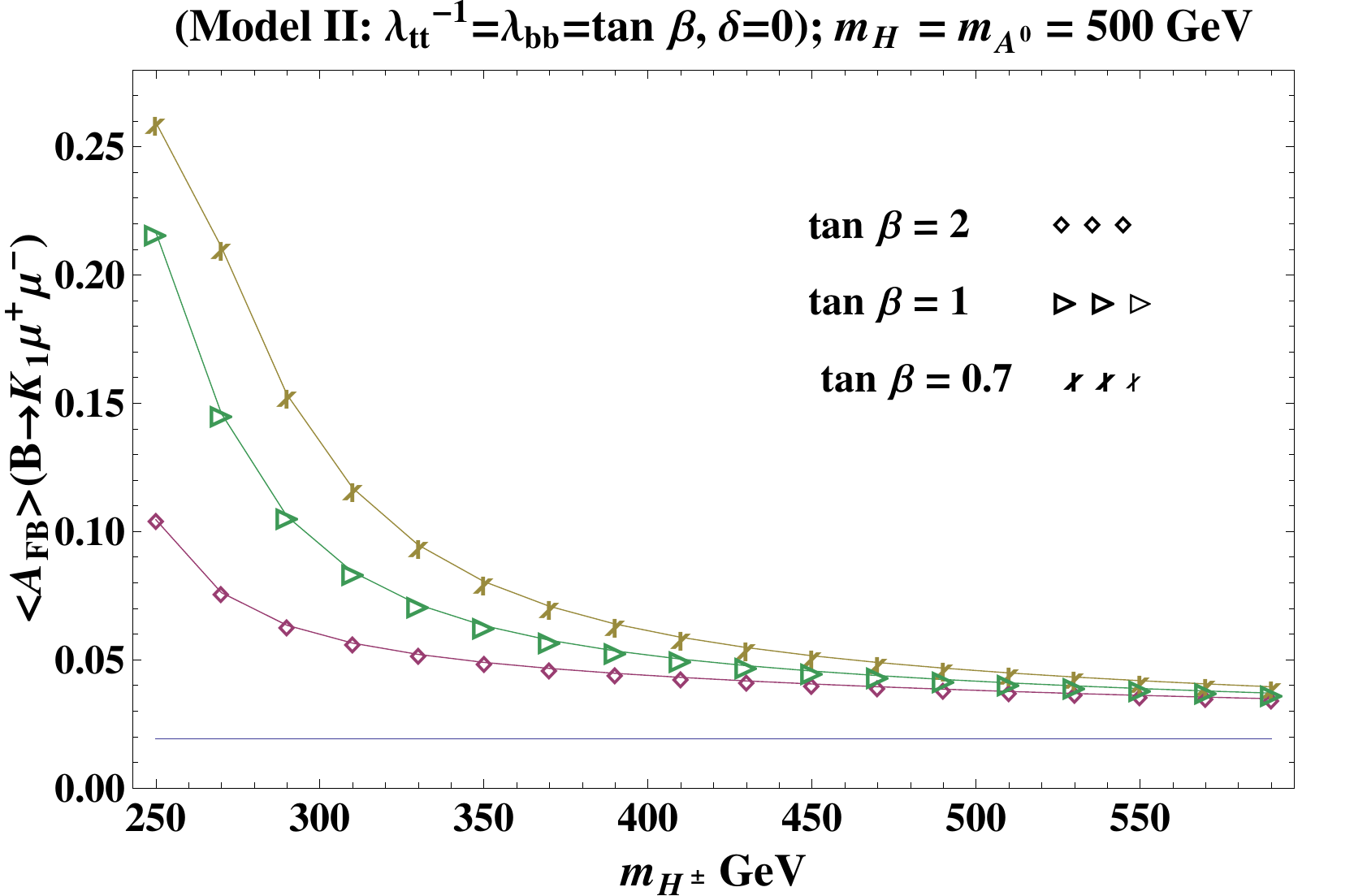}&\includegraphics[scale=0.5]{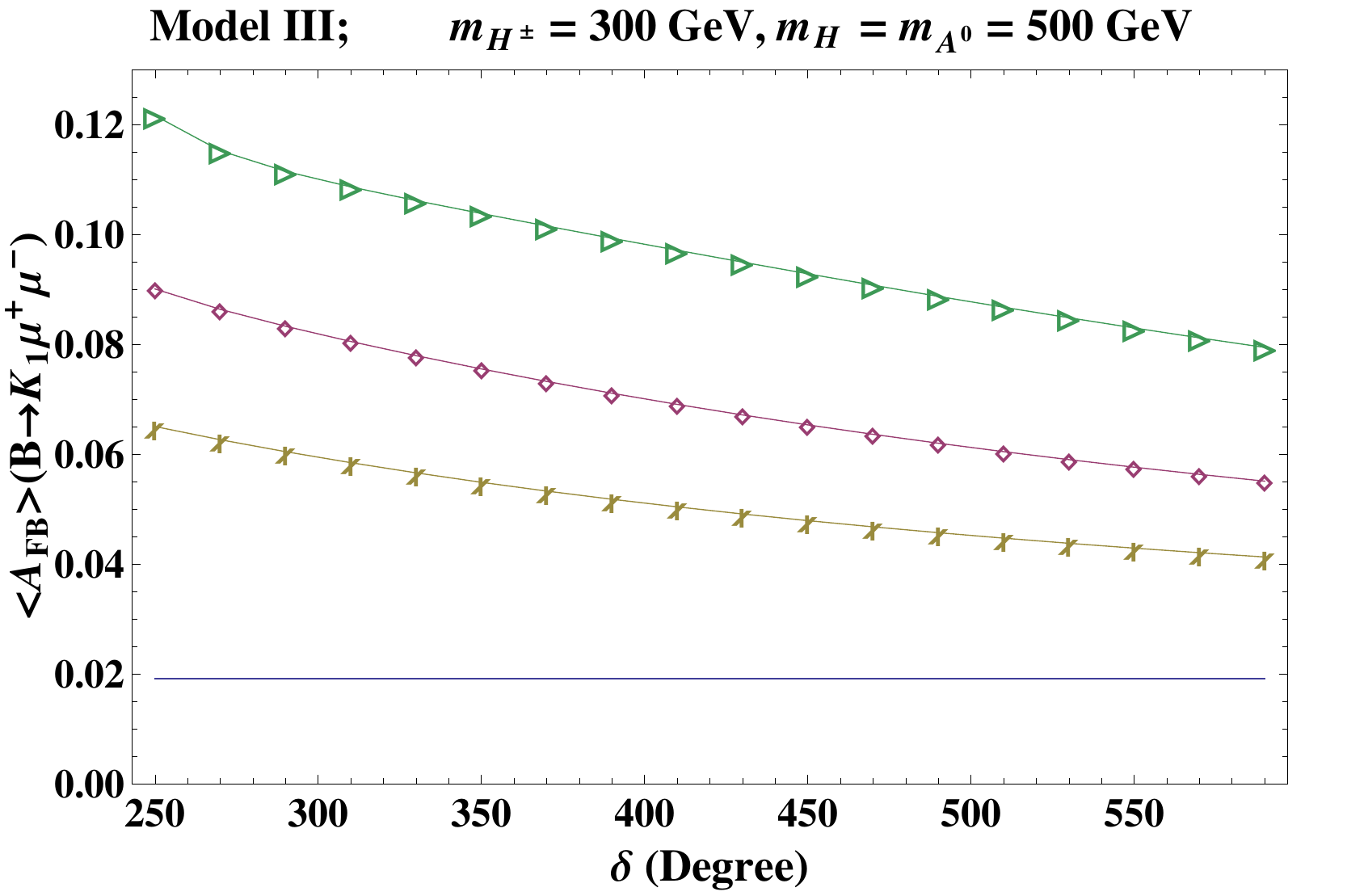}\end{tabular}
\caption{The dependence of $\langle \mathcal{A}_{FB} \rangle$ on $m_{H^\pm}$ for different values of the $\tan \beta$ in left panel and on $\delta$ in the right panel for $B \to K_1 \mu^+ \mu^-$ decay in THDM of type II and III, respectively. The values of the other parameters are given on top of each panel.}  \label{brm3}
\end{figure}

It is worth emphasizing that in addition to the NP imprint coming through the Wilson coefficients $C_7, C_9, C_{10}$ in $\mathcal{A}_{FB}$, there is also a contribution of the NP arising due to the neutral Higgs boson (NHB) coming through the auxiliary function $f_8$. It is indeed suppressed compared to the contributions from $C_7, C_9, C_{10}$ and hence its effects are too mild in the FB asymmetry.

It has already been pointed out that in $B \to K^* \ell^+ \ell^-$ the charm-loop pollution significantly modify the results of various asymmetries in different bins of the square of momentum $s$. The perturbative charm-loop contribution is usually absorbed into the definition of $C_9^{eff}$ \cite{Beneke}. The long-distance contribution is difficult to estimate,
and to incorporate them a universal correction to $C_9$ arising from the long-distance charm-loop contribution, that we parametrize as \cite{Mannel, Virto}: 
\begin{equation}
\delta C^{c \bar{c},LD}_{9} =\delta_i \frac{a+bs(c-s)}{s(c-s)} \label{cloop}
\end{equation}
with $a \in [2,7]$ GeV$^4$, $b \in [0.1,0.2]$ and $c \in [9.2, 9.5]$ GeV$^2$, where as the range of the parameter $\delta_i$ is $[-1,1]$. Because of the lack of the experimental data on the decay under consideration, the purpose here is not to scan the $A_{FB}$ in different bins of $s$ but is to see how much is the deviation by varying the parameters given in the range above. Being bold in giving the possible estimate of deviations in the value of $A_{FB}$ without charm quark loop in the SM, we have chosen $a = 3$ GeV$^{4}$, $b=0.15$, $c=9.4$ GeV$^2$ and took the value of $\delta_i = 1$ (purple curve) and $\delta_i = -1$ (golden curve) in Fig. \ref{cloop-figure}. We can see that the maximum shift in the value of forward-backward asymmetry is around $20 \%$ from the case when charm-loop pollution is ignored. In Figs. \ref{brm1} and \ref{brm2} we can see that in certain range of the parameters of THDM, the deviation from the SM value is significantly large. Therefore, in future, when we have data on these decays, it will be possible to limit the parametric space of THDM as well as of the parameters corresponding to charm-loop effects.

Besides the zero position of $\mathcal{A}_{FB}$, its magnitude will also serve as an important tool to see the imprints of NP. The average value of $\mathcal{A}_{FB}$, after integration on $s$ in the range which is below the resonances, i.e., $4m^2_{\ell} \leq s \leq 9 GeV^2$ for $B \to K_1 \mu^+ \mu^-$ is displayed in Fig. \ref{brm3}. In Fig. \ref{brm3}(a) the variations of $\langle \mathcal{A}_{FB} \rangle$ with the mass of charged Higgs boson $(m_{H^\pm})$ for different values of $\tan \beta$ is portrayed. It can observed that for the small value of the mass of charged Higgs boson $(m_{H^\pm})$ the $\langle \mathcal{A}_{FB} \rangle$ significantly enhanced by enhancing the value of $\tan \beta$ in $B \to K_1 \mu^+ \mu^-$ decay (c.f. Fig. \ref{brm3}(a). However, this value become less sensitive to the value of $\tan \beta$ at large value of $m_{H^\pm}$. This is because of the fact that  the value of parameter $y=\frac{m^2_t}{m^2_{H^\pm}}$ decreases and so the corresponding NP effects become small. 

Likewise, we have also shown the dependence of the $\langle \mathcal{A}_{FB} \rangle$ on the $CP$ violating phase $\delta$ arises in model III in Fig. \ref{brm3}(b). Here, we have kept the the mass of charged Higgs to be 300 GeV and varied the values of $\lambda_{bb}$ and $\lambda_{tt}$ for different cases defined in Eq. (\ref{lambdas}). It can be seen that for small value of the phase, the increase in the value of $\lambda_{tt}$ will lead to increase in the value of the magnitude of $\mathcal{A}_{FB}$ and at the phase value to be $90^\circ$, this value for all the three cases become same. Hence, being insensitive to the uncertainties arising due to different input parameters, the deviations in the magnitude of $\mathcal{A}_{FB}$ due to the THDM parameters are very prominent and easy to measure at the experiment which can also help us to put constraints on the parameter space of different versions of THDM.

\subsection{Analysis of Lepton Polarization Asymmetries}
\begin{figure}[ht]
\begin{tabular}{cc}
\centering
\hspace{0.5cm}($\mathbf{a}$)&\hspace{1.2cm}($\mathbf{b}$)\\
\includegraphics[scale=0.6]{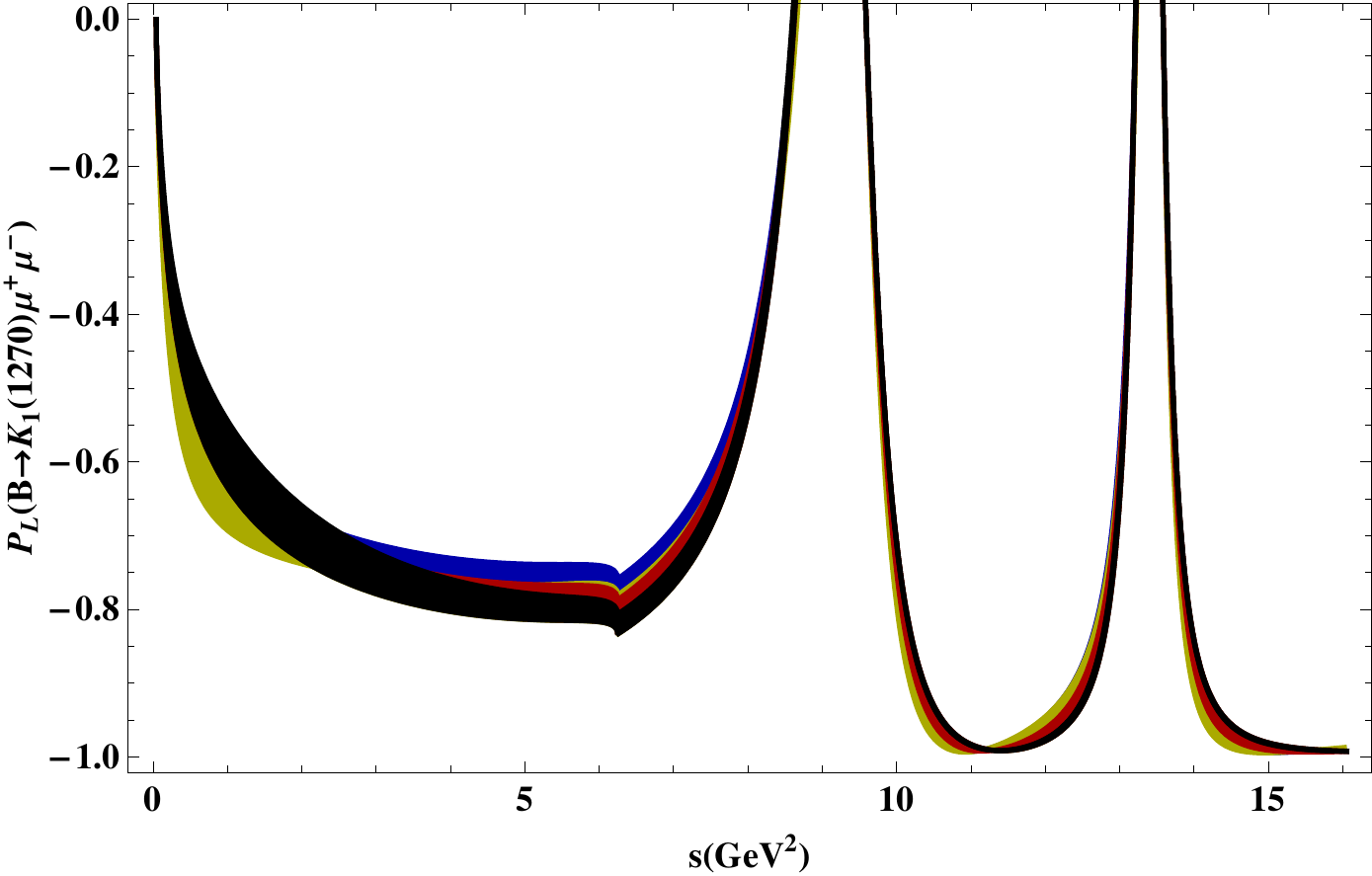}&\includegraphics[scale=0.6]{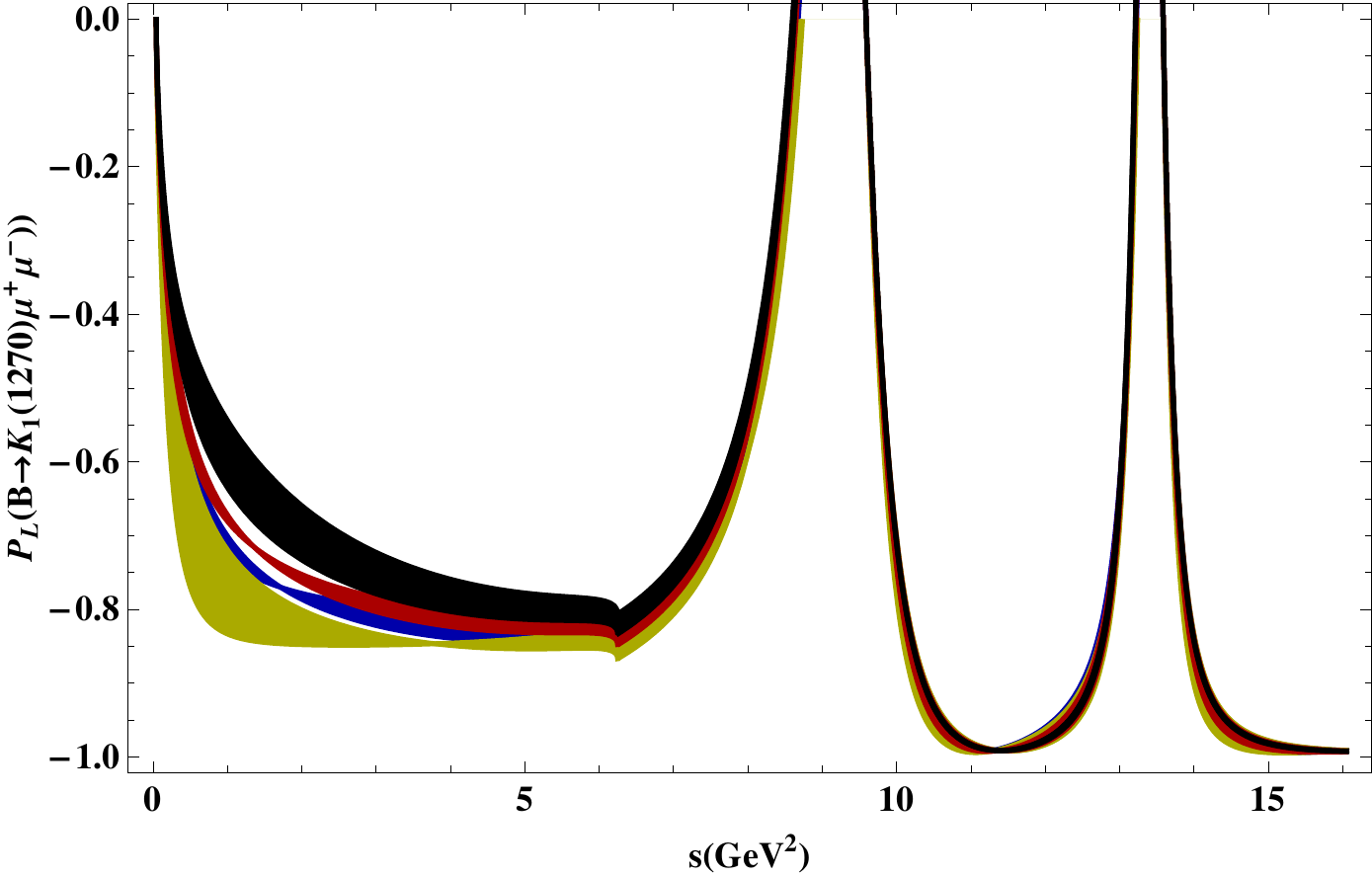}\\
\hspace{0.5cm}($\mathbf{c}$)&\hspace{1.2cm}($\mathbf{d}$)\\
\includegraphics[scale=0.6]{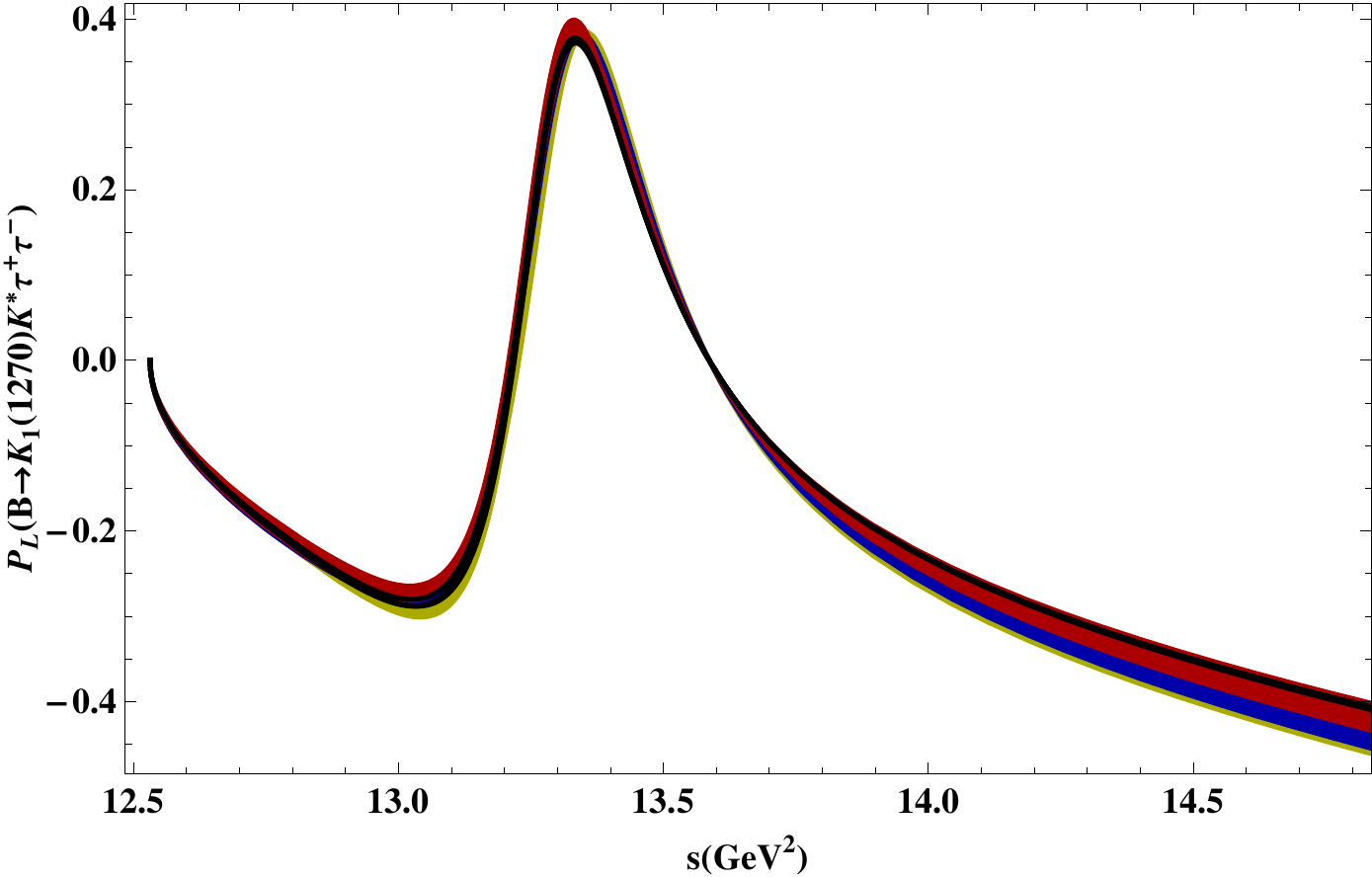}&\includegraphics[scale=0.6]{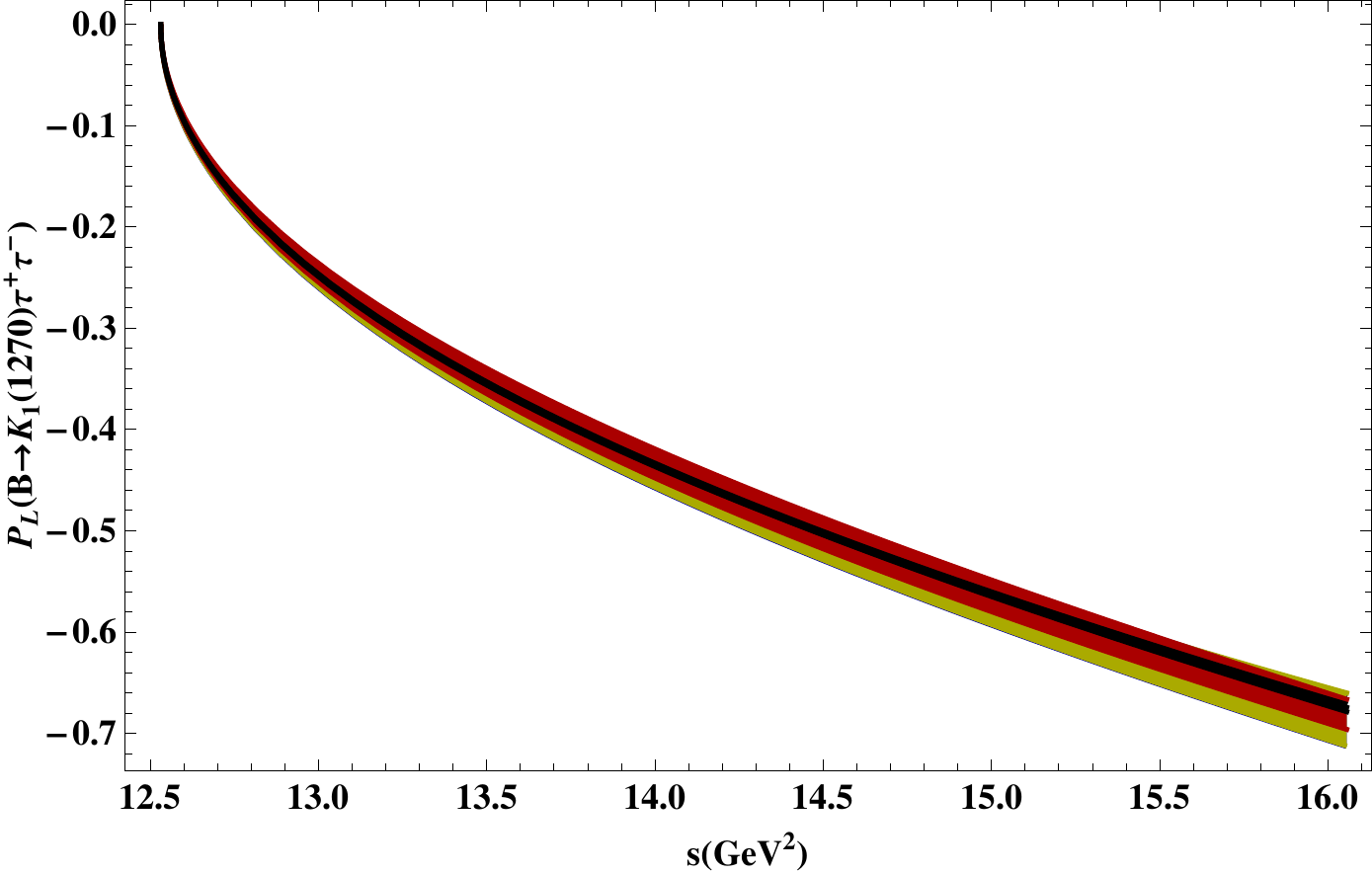}\end{tabular}
\caption{The dependence of longitudinal polarization asymmetries of $B\to K_{1}\ell^{+}\ell^{-}$ on $s$ with long-distance contributions for (a) muons, (c) tauons in THDM1 and for (b) muons (d) tauons in THDM2 where bands are shown the range of $\tan\beta$ from 1 to 30 degrees. In all the graphs black band corresponds to the SM and yellow, blue and red bands correspond to the values of $m_H^{\pm}=300$GeV, $m_H^{\pm}=400$GeV and $m_H^{\pm}=700$GeV, respectively, while the values of $m_H^0$ and $m_A$ are set to be $500$GeV.} \label{lp1}
\end{figure}

\begin{figure}[ht]
\begin{tabular}{cc}
\centering
\hspace{0.5cm}($\mathbf{a}$)&\hspace{1.2cm}($\mathbf{b}$)\\
\includegraphics[scale=0.6]{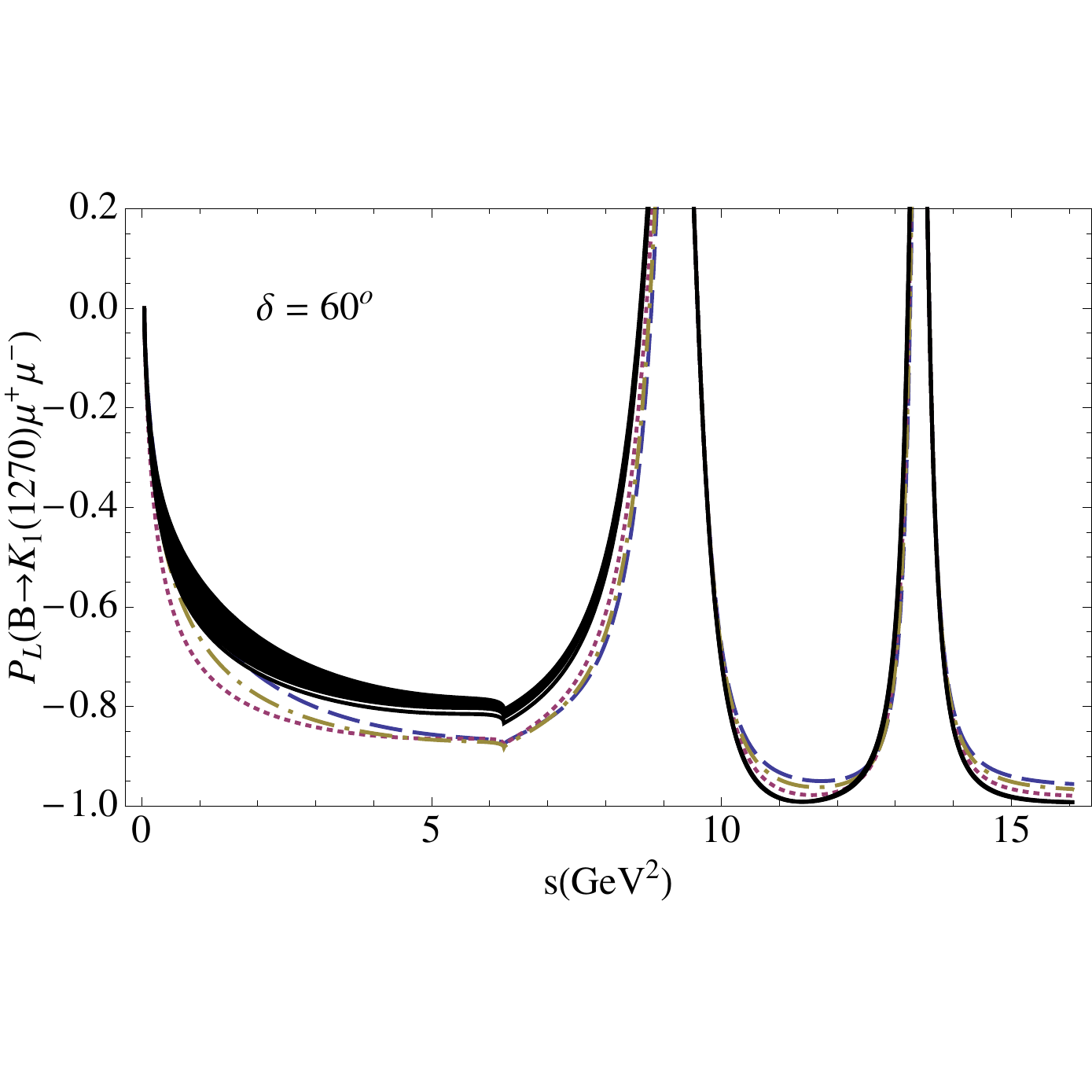}&\includegraphics[scale=0.6]{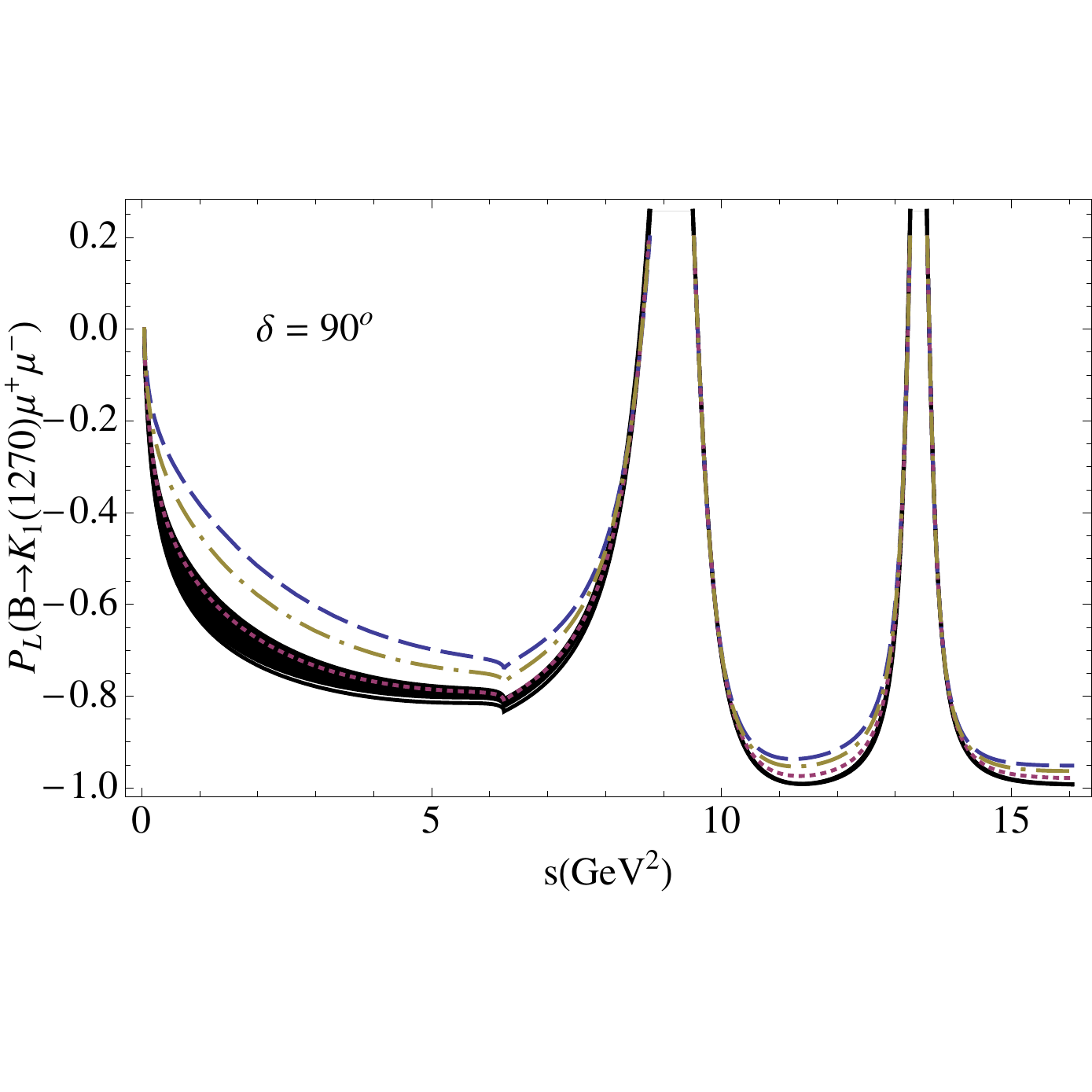}\\
\hspace{0.5cm}($\mathbf{c}$)&\hspace{1.2cm}($\mathbf{d}$)\\
\includegraphics[scale=0.6]{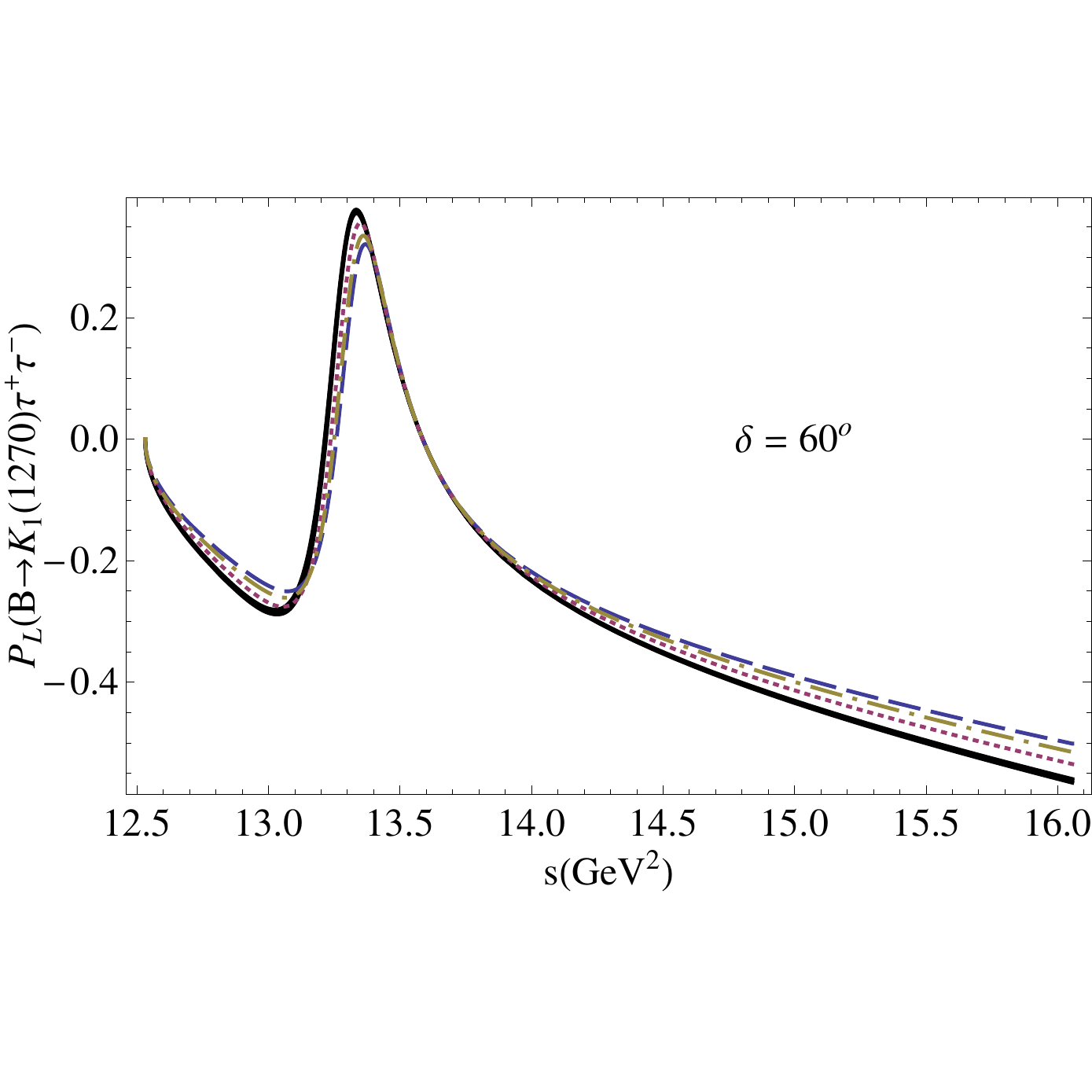}&\includegraphics[scale=0.6]{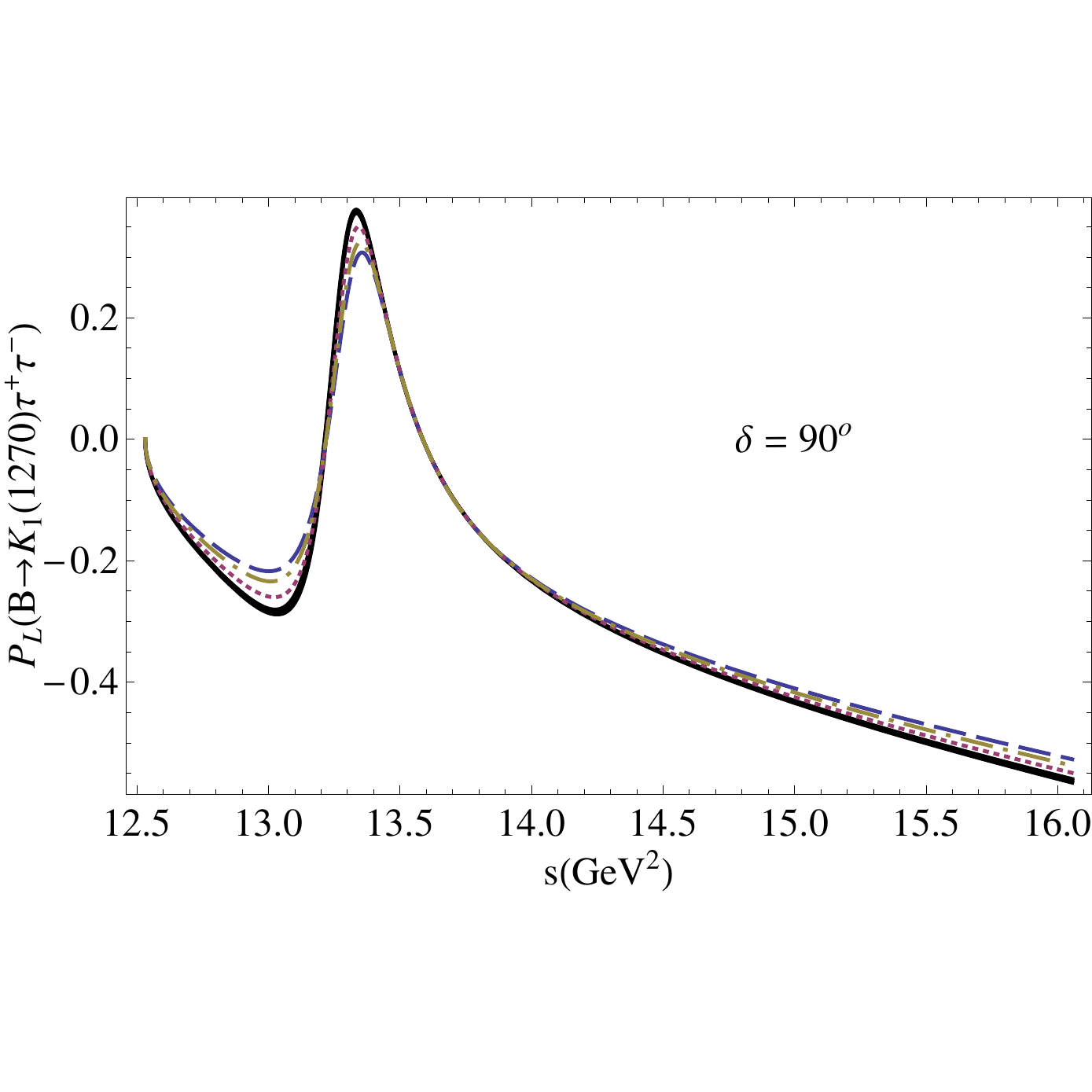}\end{tabular}
\caption{The dependence of longitudinal polarization asymmetries of $B\to K_{1}\ell^{+}\ell^{-}$ on $s$ with long-distance contributions for muons (a) $\delta=60^\circ$ (b) $\delta=90^\circ$  and for tauons (c) $\delta=60^\circ$ (d) $\delta=90^\circ$ in THDM3. The line convention as well as the values of parameters corresponding to the version III of THDM is same as in Fig. \ref{brm2}} \label{lp2}
\end{figure}

\begin{figure}[ht]
\begin{tabular}{cc}
\centering
\hspace{0.5cm}($\mathbf{a}$)&\hspace{1.2cm}($\mathbf{b}$)\\
\includegraphics[scale=0.5]{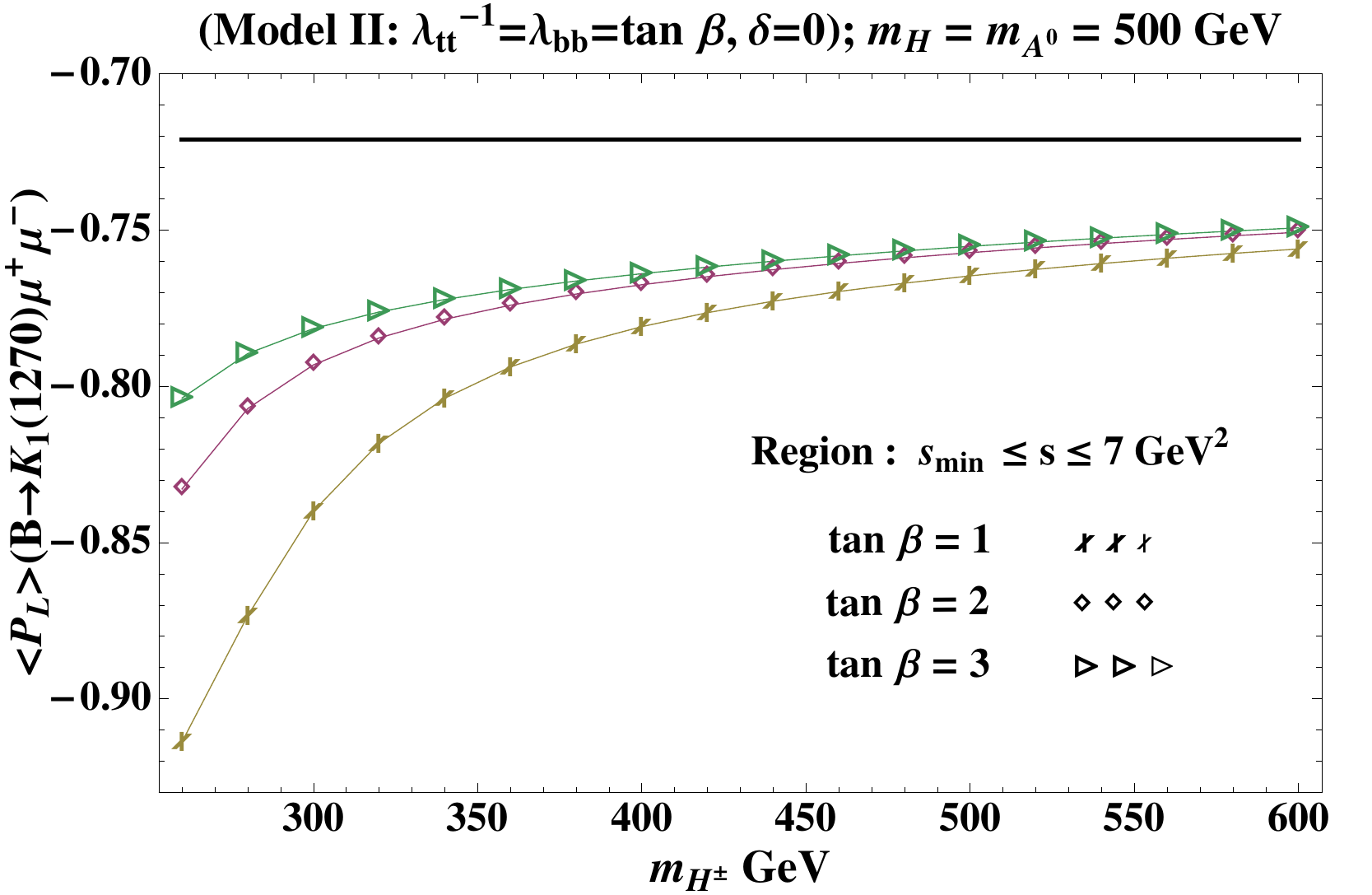}&\includegraphics[scale=0.5]{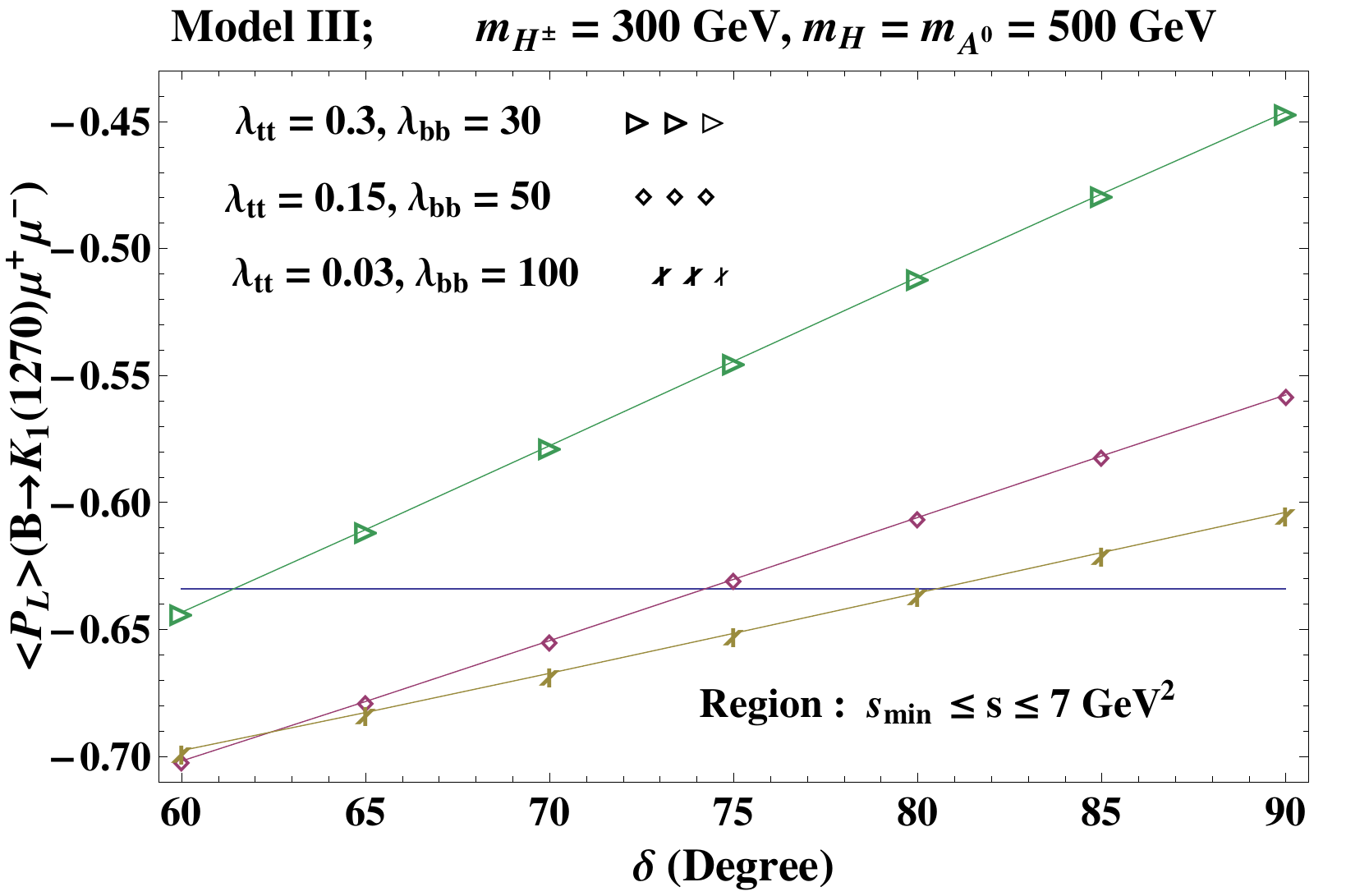}\end{tabular}
\caption{The dependence of $\langle P_{L} \rangle$ on $m_{H^\pm}$ for different values of the $\tan \beta$ in left panel and on $\delta$ in the right panel for $B \to K_1 \mu^+ \mu^-$ decay in THDM of type II and III, respectively. The values of the other parameters is given on top of each panel.} \label{lp3}
\end{figure}

\begin{figure}[ht]
\begin{tabular}{cc}
\centering
\hspace{0.5cm}($\mathbf{a}$)&\hspace{1.2cm}($\mathbf{b}$)\\
\includegraphics[scale=0.6]{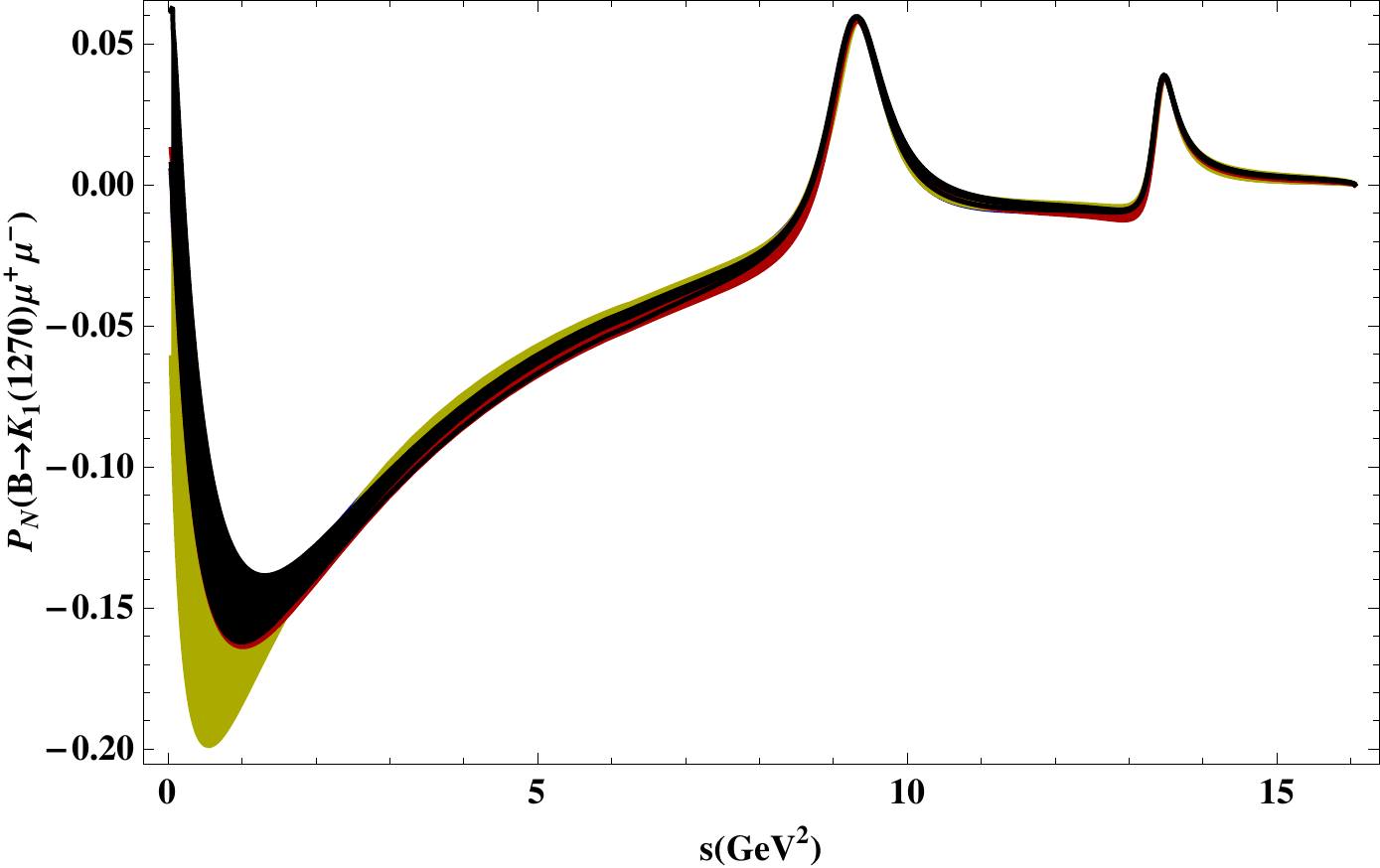}&\includegraphics[scale=0.6]{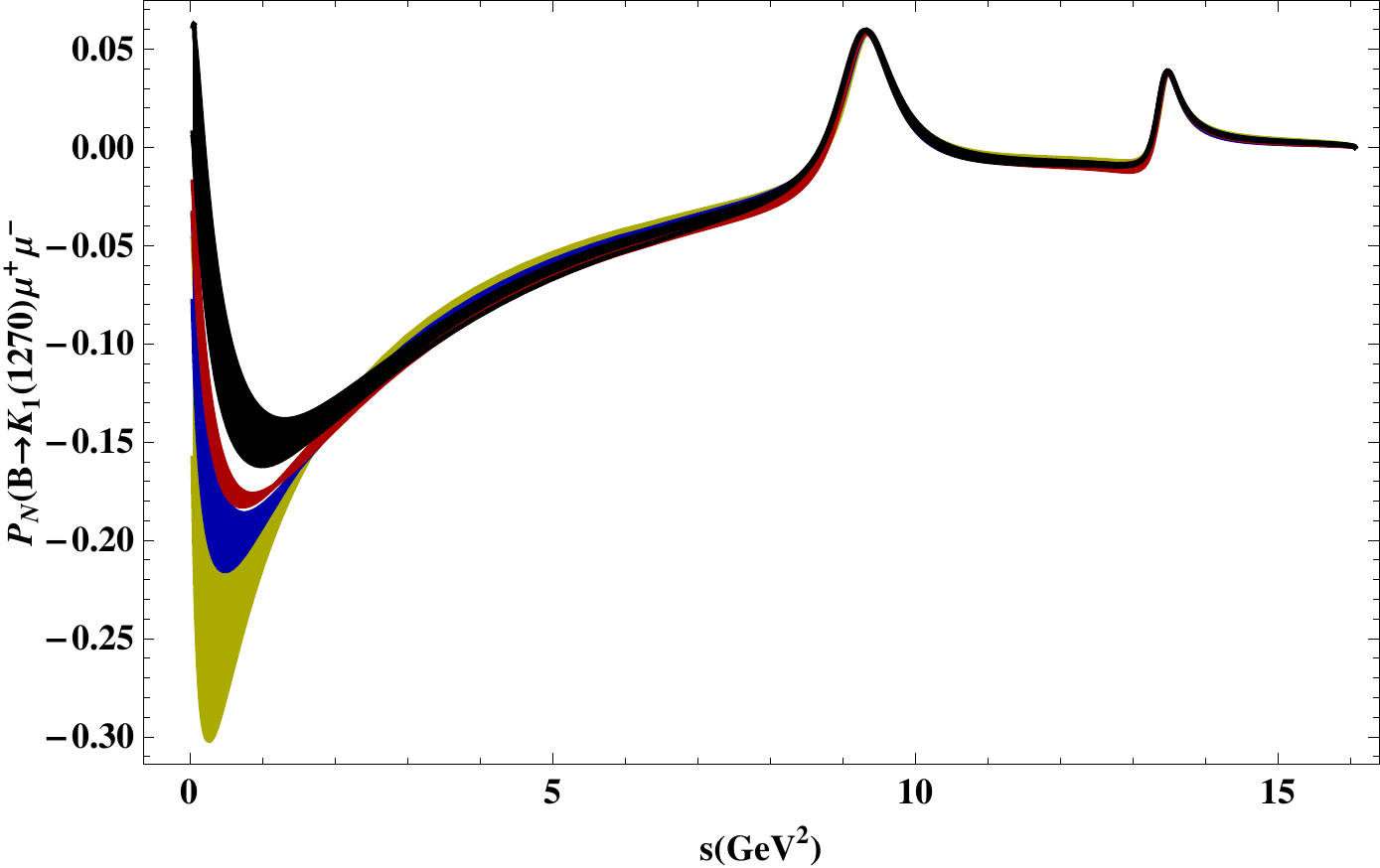}\\
\hspace{0.5cm}($\mathbf{c}$)&\hspace{1.2cm}($\mathbf{d}$)\\
\includegraphics[scale=0.6]{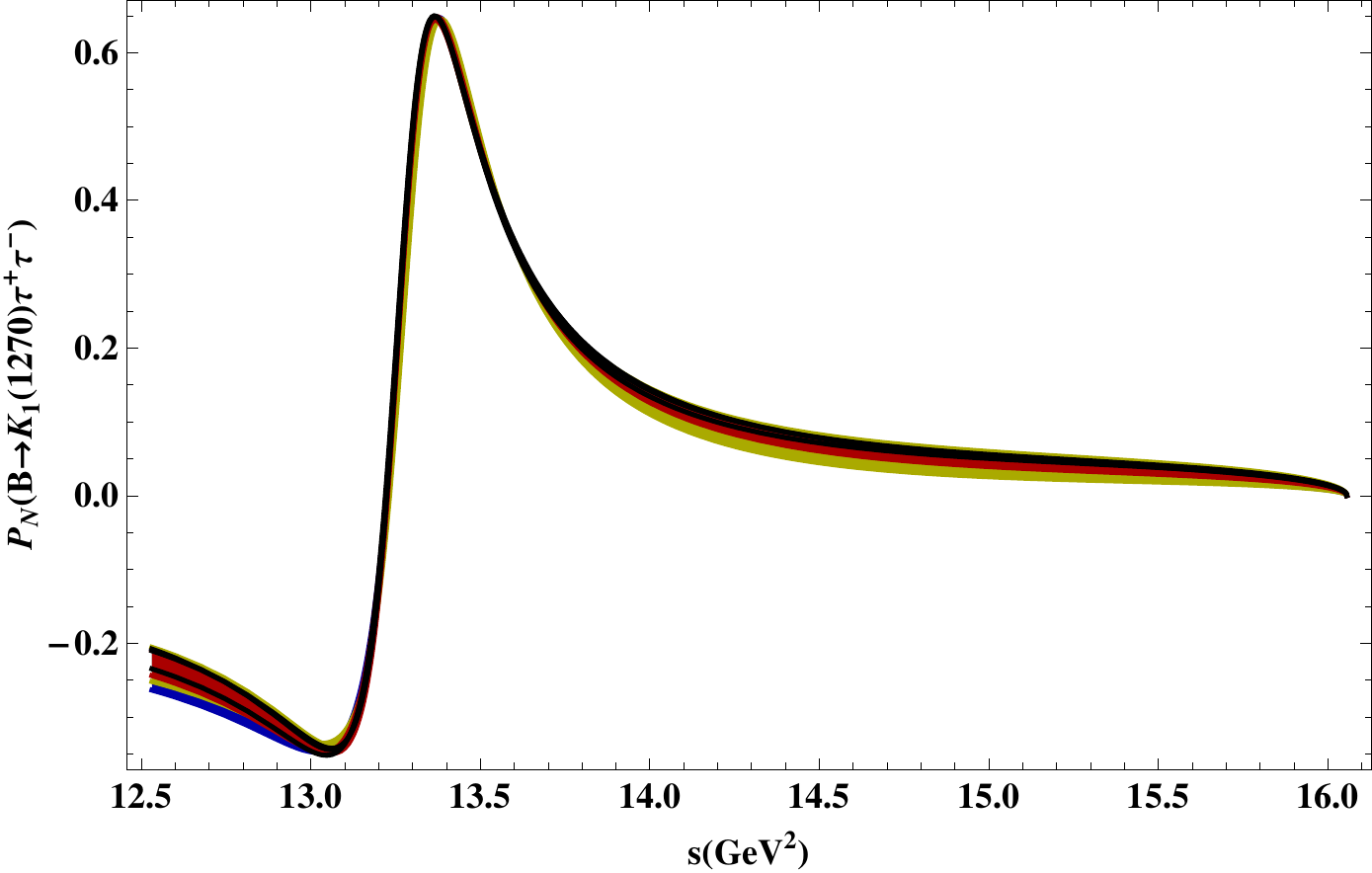}&\includegraphics[scale=0.6]{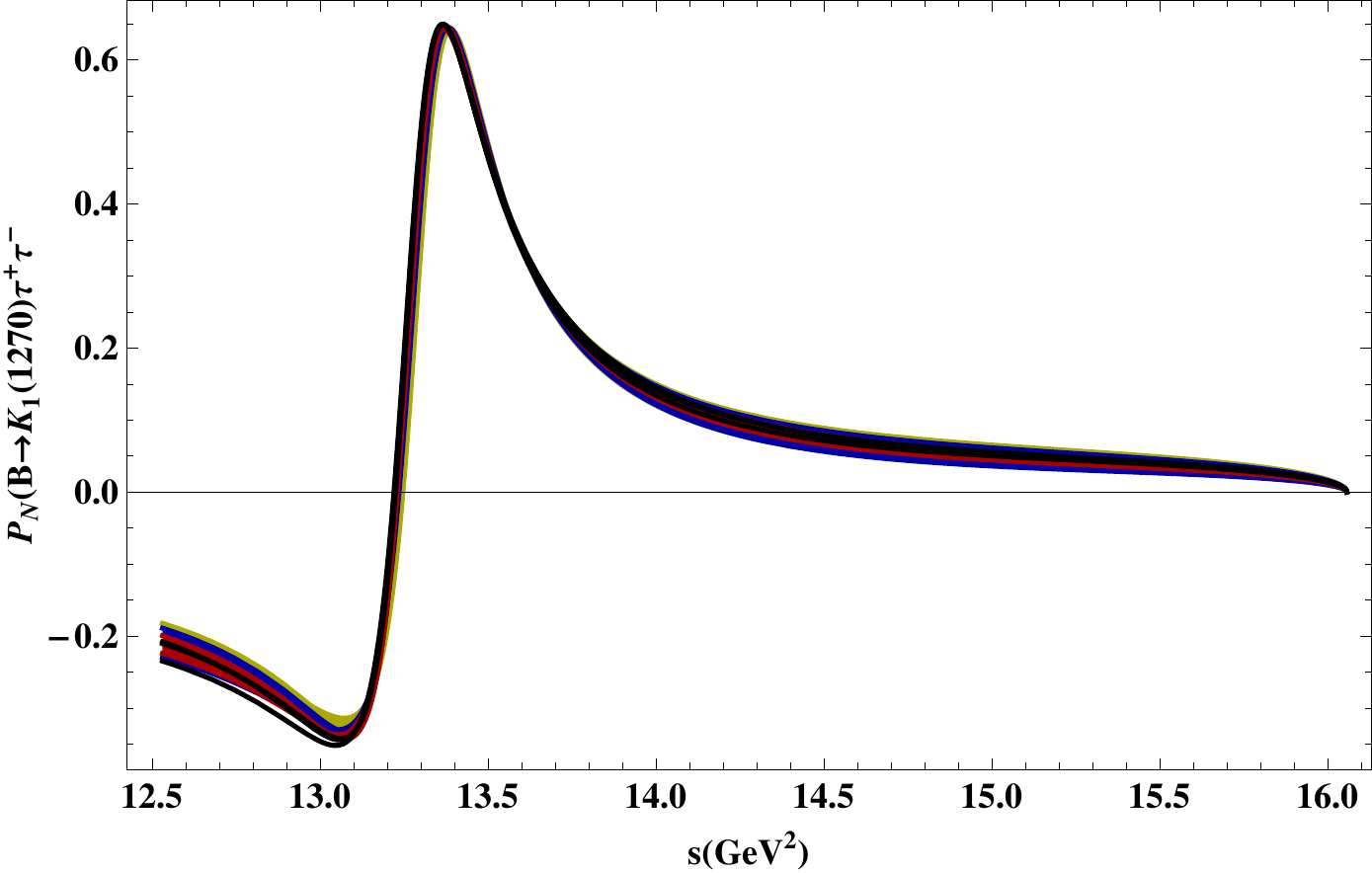}\end{tabular}
\caption{The dependence of normal polarization asymmetries of $B\to K_{1}\ell^{+}\ell^{-}$ on $s$ with long-distance contributions for (a) muons, (c) tauons in THDM1 and for (b) muons (d) tauons in THDM2 where bands are shown the range of $\tan\beta$ from 1 to 30 degrees. In all the graphs black band corresponds to the SM and yellow, blue and red bands correspond to the values of $m_H^{\pm}=300$GeV, $m_H^{\pm}=400$GeV and $m_H^{\pm}=700$GeV, respectively, while the values of $m_H^0$ and $m_A$ are set to be $500$GeV.} \label{pn1}
\end{figure}

\begin{figure}[ht]
\begin{tabular}{cc}
\centering
\hspace{0.5cm}($\mathbf{a}$)&\hspace{1.2cm}($\mathbf{b}$)\\
\includegraphics[scale=0.6]{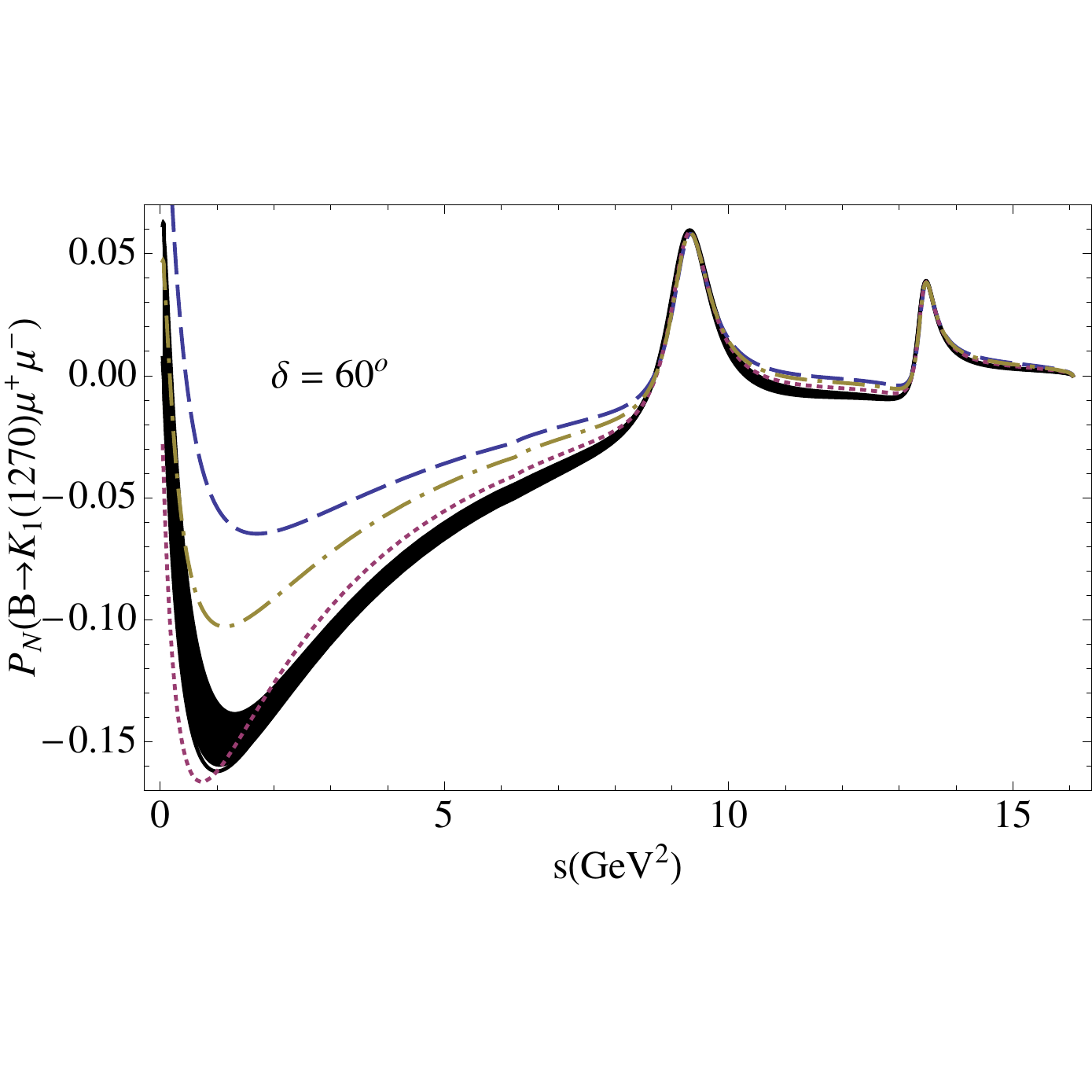}&\includegraphics[scale=0.6]{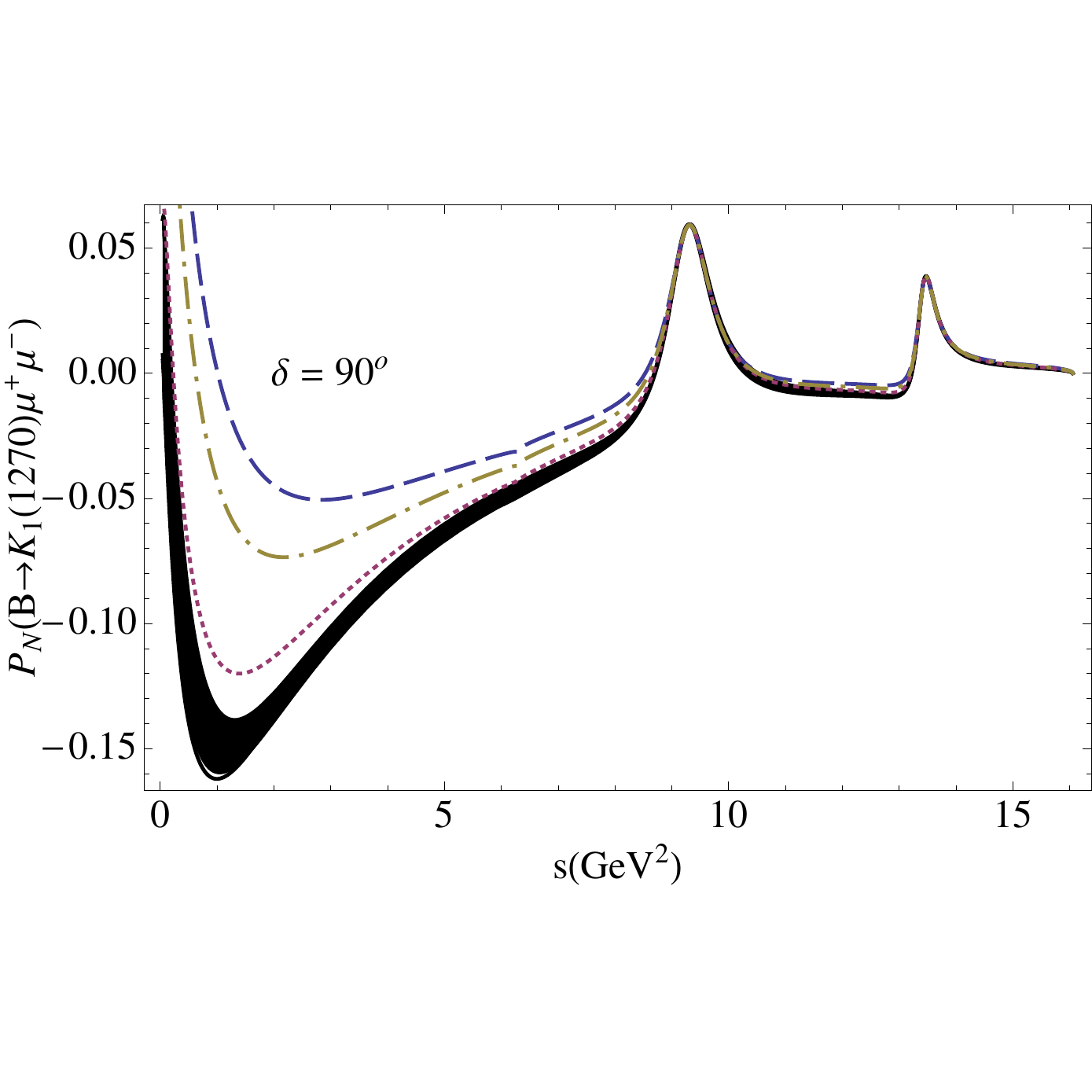}\\
\hspace{0.5cm}($\mathbf{c}$)&\hspace{1.2cm}($\mathbf{d}$)\\
\includegraphics[scale=0.6]{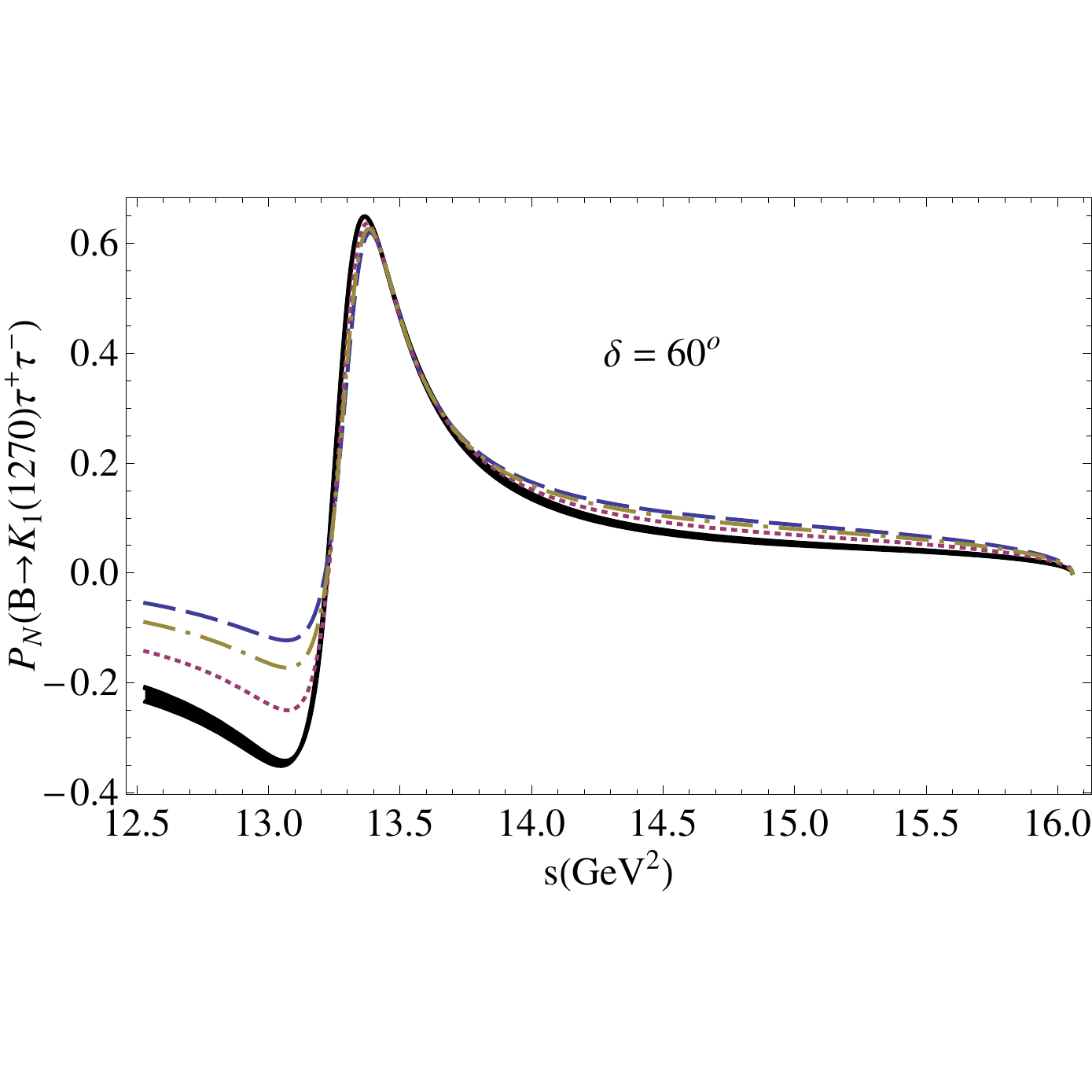}&\includegraphics[scale=0.6]{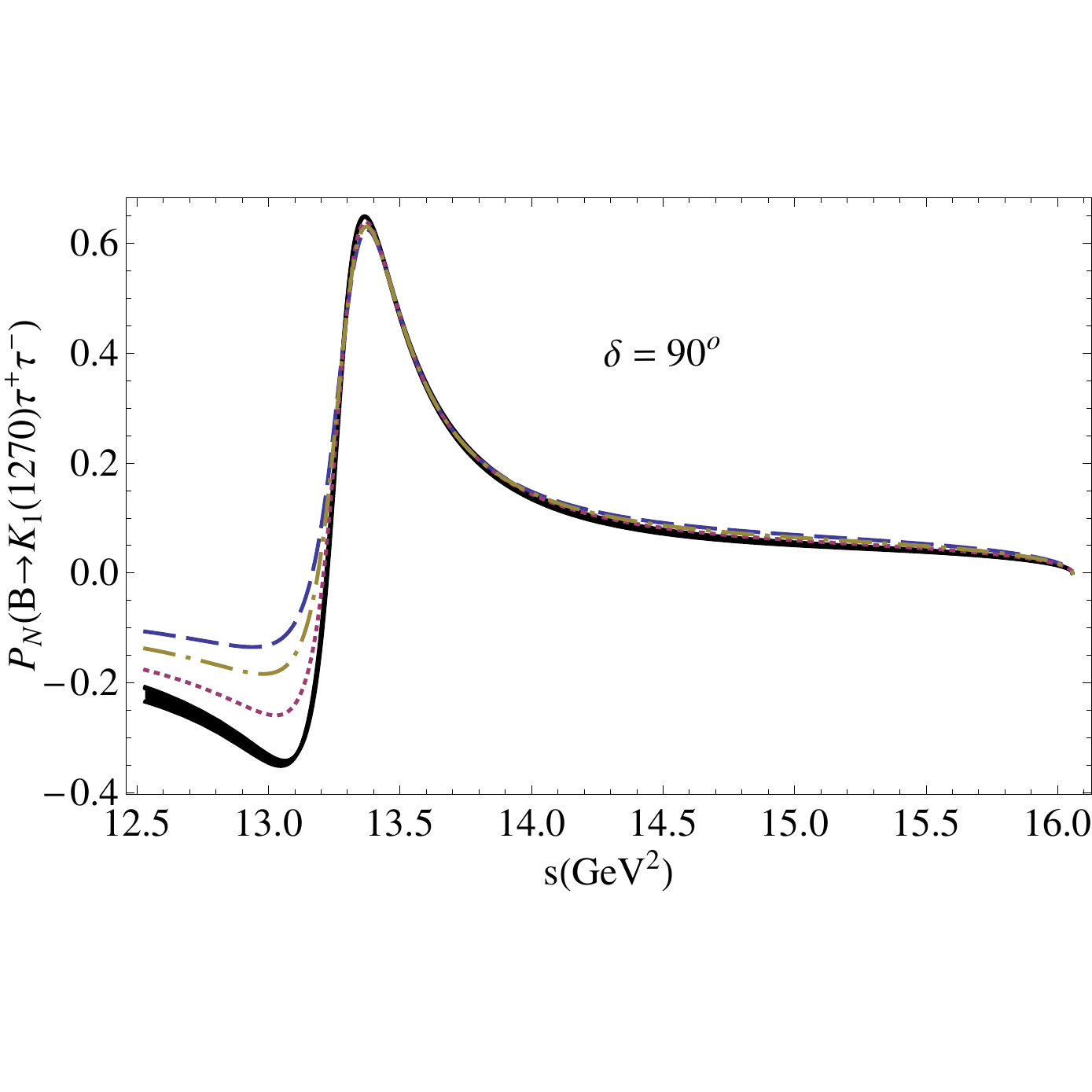}\end{tabular}
\caption{The dependence of normal polarization asymmetries of $B\to K_{1}\ell^{+}\ell^{-}$ on $s$ with long-distance contributions for muons (a) $\delta=60^\circ$ (b) $\delta=90^\circ$  and for tauons (c) $\delta=60^\circ$ (d) $\delta=90^\circ$ in THDM of type III. In all the graphs solid band corresponds to the SM uncertainties. The dashed, dotted and dashed-dot lines correspond Case A, Case B and Case C (c.f. Eq. (\ref{lamdavalues})), respectively.} \label{pn2}
\end{figure}

\begin{figure}[ht]
\begin{tabular}{cc}
\centering
\hspace{0.5cm}($\mathbf{a}$)&\hspace{1.2cm}($\mathbf{b}$)\\
\includegraphics[scale=0.5]{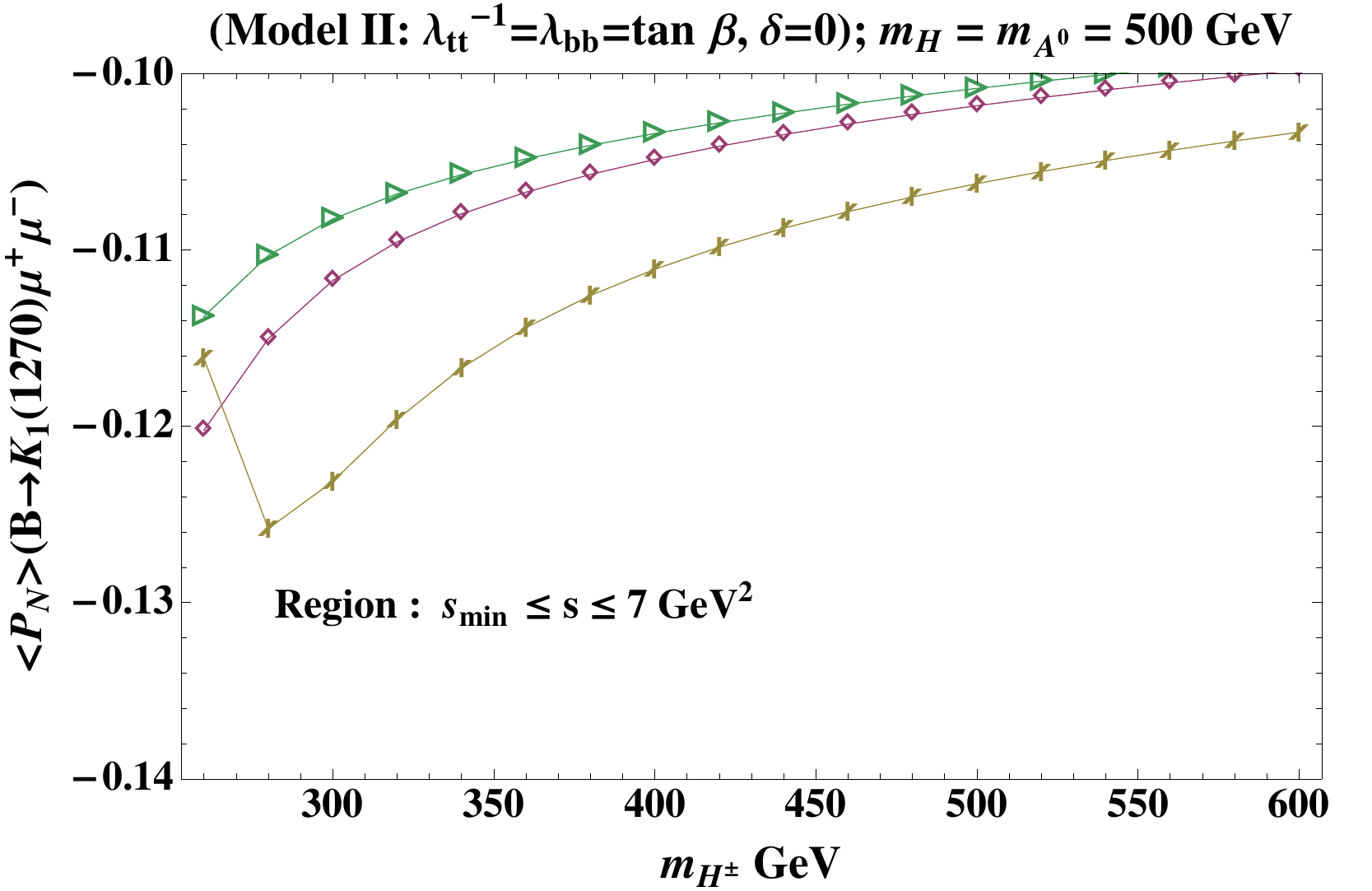}&\includegraphics[scale=0.5]{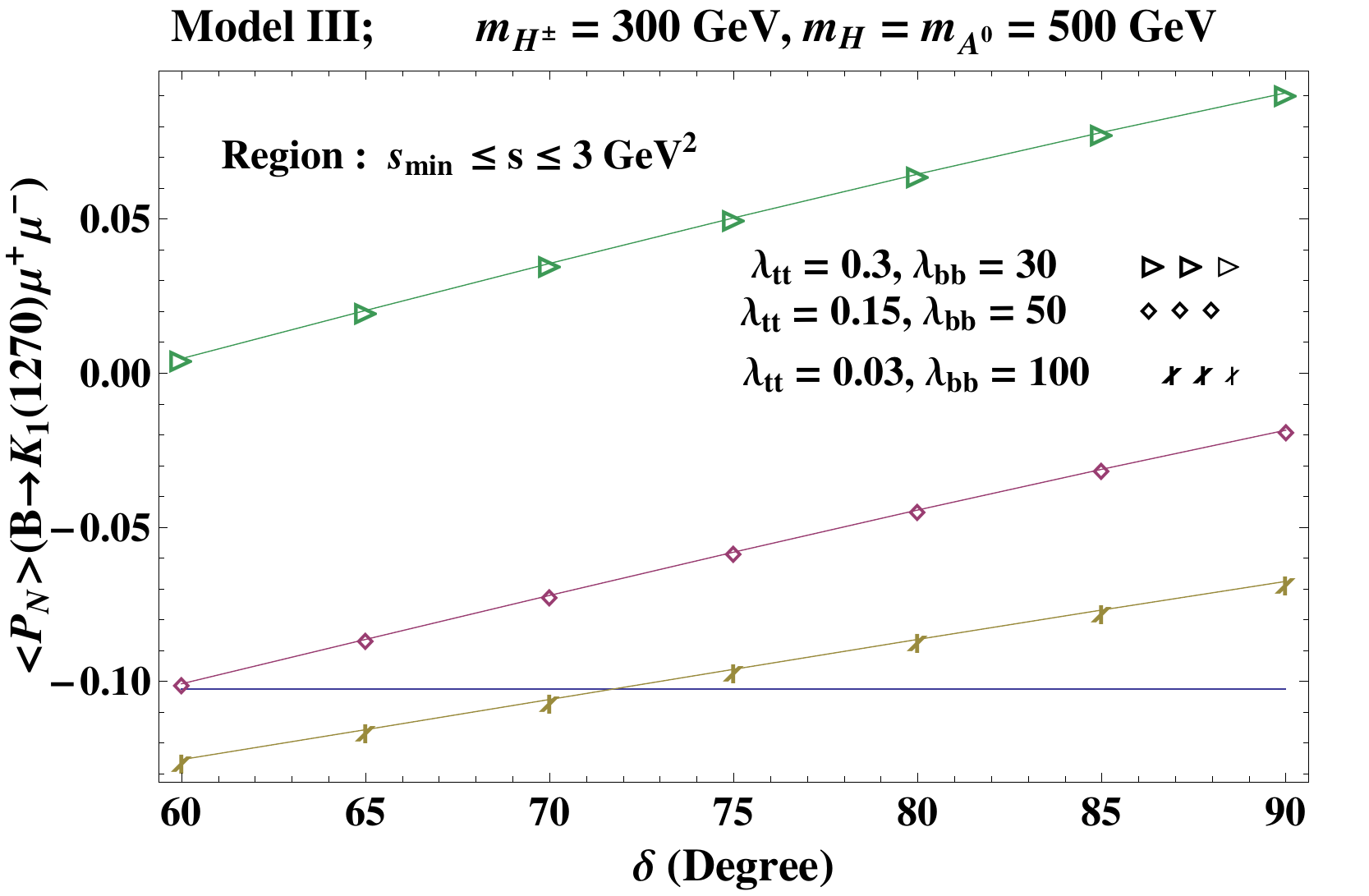}\end{tabular}
\caption{The dependence of $\langle P_{N} \rangle$ on $m_{H^\pm}$ for different values of the $\tan \beta$ in left panel and on $\delta$ in the right panel for $B \to K_1 \mu^+ \mu^-$ decay in THDM of type II and III, respectively. The lines with cross, box and triangle correspond to the values of  $\tan \beta$ equal to 1, 2 and 3, respectively. The values of the other parameters is given on top of each panel.} \label{pn3}
\end{figure}

In addition to the forward-backward asymmetry, the other interesting asymmetries to get the complementary information about NP associated with the THDM in $B\to K_{1} \ell^{+}\ell^{-}$ $(\ell=\mu,\tau)$ decays, are the lepton polarization asymmetries which are shown in Figs. \ref{lp1}, \ref{lp2}, \ref{lp3}, \ref{pn1}, \ref{pn2} and \ref{pn3}. Since, lepton polarization asymmetries depend on the different combinations of the Wilson Coefficients, therefore, one can expect large dependency of these asymmetries on different versions of THDM and hence making these observables fertile to hunt the possible NP. In case of the longitudinal lepton polarization $(P_L)$, it can be seen from Eq. (\ref{long-polarization}) that the contribution from the NHB encoded in the Wilson coefficients $C_{Q_1}$ and $C_{Q_2}$ is suppressed by the mass of final state leptons. In addition, these coefficients have a factor of $|\lambda_{tt}|^2$ in the denominator (c.f. Eqs. (\ref{cq1}) and (\ref{cq2})) and in model III, where this factor is less than 1, it lifts the suppression due to the mass of lepton. Therefore, one can expect the large contribution from NHB in THDM of type III.  Figs. \ref{lp1}(a) and \ref{lp1}(b) display the trend of $P_{L}$ with the square of momentum transfer in $B \to K_1 \mu^+ \mu^-$ decay. It can be seen that even for small values of the mass of charged Higgs boson, $m_{H^\pm}$, and also because of the like sign of $\lambda_{tt}$ and $\lambda_{bb}$ the signature of NP coming through THDM type II are prominent at small value of $s$ (c.f. Fig. \ref{lp1}(b)). However, in case of $\tau$'s as final state leptons, the NP effects overlap with each other and appear only at high value of $s$ because of the pre-factor of $(s-\frac{4m^2_{\ell}}{s})$ in Eq. (\ref{long-polarization}) and it is evident from Figs. \ref{lp1}c and \ref{lp1}d for THDM of types I and II, respectively.

Contrary to the types I and II, in version III of THDM one can expect a large contribution from the NHBs because of the proportionality of inverse of $|\lambda_{tt}|^2$  that would lift the lepton mass suppression coming in the last term of longitudinal lepton polarization asymmetry. As $|\lambda_{tt}|$ is less than one in type III, therefore, the terms proportional to the square of  $|\lambda_{tt}|$ are ignorable compared to the terms linear in $\lambda_{tt}\lambda_{bb}$ and hence only the THDM type III contributions will be prominent (c.f. Eq. (\ref{C7mw})). As the term $\lambda_{tt}\lambda_{bb}$ contains phase $\delta$, therefore, $P_{L}$ will also be sensitive to the phase $\delta$ and this can be witnessed from Fig. \ref{lp2}. In case of the muons as final state leptons, one can see from Figs. \ref{lp2}(a), \ref{lp2}(b) that at $\delta = 60^{\circ}$ the contribution from the model III will lead to contribution which make the value of the $P_{L}$ compared to the SM value and also it lies away from the uncertainty region. However, the trend is entirely different in case of $\delta = 90^{\circ}$. Compared to muons, when the final state leptons are $\tau$'s the effects of the type III gives positive contribution for both values of the phase. However, in this case the effects are mild but still distinguishable from the SM. 

In order to make the impact of NP coming through THDM, we have plotted the average value of the longitudinal lepton polarization asymmetry $\langle P_L \rangle$ with the charged Higgs boson mass (Fig. \ref{lp3}(a)) and with $CP$ violating phase $\delta$ (Fig. \ref{lp3}(b)) in versions II and III of the THDM, respectively. The integration on square of momentum is performed in the range $s_{min} \leq s \leq 7 GeV^2$, i.e., well below the resonance region. Fig. \ref{lp3}a shows that for certain values of the parameters in THDM the shift in $\langle P_L \rangle$ is significant at small value of the mass of charged Higgs boson. However, this shift in $\langle P_L \rangle$ diminish at the large value of $m_{H^\pm}$ due to the fact that NP entering in the Wilson coefficients is partly through the parameter $y=m^2_{t}/m^2_{H^\pm}$ which becomes small at large value of $m_{H^\pm}$. Similarly, Fig. \ref{lp3}b depicts the behaviour of $\langle P_{L}\rangle$ with the phase $\delta$ for different values of $\lambda_{tt}$ and $\lambda_{bb}$. It can be observed that for large value of $\delta$ along with $\lambda_{tt}=0.3$ and $\lambda_{bb}=30$, the average value of $P_L$ is significantly modified which is likely to be measured at some of the on going and future experiments.

Figs. \ref{pn1} shows the trend of normal lepton polarization asymmetry $(P_N)$ in $B \to K_1 \ell^+ \ell^-$ decay in versions I and II of the THDM. It can be seen that the most prominent change in the value of the $P_N$ for $B \to K_1 \mu^+ \mu^-$ decay comes at small value of the mass of the charged Higgs boson which is shown by the yellow band. However, at large value of the mass of Higgs boson the variations due to $\tan \beta$ are small and also the value approaches to the SM results. It is because of the fact that the value of the variable $y$ decreases due to increase in the value of $m_{H^{\pm}}$ and so the NP content becomes small in the Wilson coefficients. 

Figures \ref{pn2}(a,b) and \ref{pn2}(c,d) show the dependence of normal lepton polarization asymmetries with the square of momentum transfer when we have $\mu$'s and $\tau$'s as final state leptons, respectively in THDM version III for different values of $CP$ violating phase $\delta$. The black solid band correspond to the SM results by including the uncertainties involved in different input parameters like the form factors, etc. One can notice that the value of $P_N$ are quite sensitive to the \textit{Case A} and \textit{C} (c.f. Eq. (\ref{lamdavalues})) which are depicted by dashed and dashed-dotted lines in Figs. \ref{pn2} for $\delta =60^\circ$ and $90^\circ$. 

To be more clear about the influence of THDM's parameters, we have plotted the average value of normal lepton polarization asymmetry, $\langle P_N \rangle$ against the mass of the charged Higgs boson $(m_{H^\pm})$ and $CP$ violating phase $\delta$ in Figs. \ref{pn3}a and \ref{pn3}b, respectively for $B \to K_1 \mu^+ \mu^-$ decay. In Fig. \ref{pn3}a, one can notice that at small value of $m_{H^\pm}$ the average value of normal lepton polarization asymmetry is very much sensitive to the value of $\tan \beta$ in THDM type II. We can see that increasing the value of $\tan \beta$ the value of $\langle P_N \rangle$ increases from $-0.105$ to $-0.125$ when the value of $m_{H^\pm}$ is fixed to $300$GeV. However, at large value of the $m_{H^\pm}$ the value is no more sensitive to the parameters of THDM of type II. Likewise, the average value of $P_N$ is also sensitive to the parameters of THDM of type III which is depicted in Fig. \ref{pn3}b. In this figure, we have ntegrated on $s$ in the range $s_{min} \leq s \leq 3 GeV^2$ because the most visible effects comes in this bin of $s$.  In Fig. \ref{pn3}b it can be noticed that $\langle P_N \rangle$ is quite sensitive to the parameters $\lambda_{tt}$, $\lambda_{bb}$ and $\delta$. We can see that value of $\langle P_N \rangle$ become more negative when $\lambda_{tt}$ is decreased from $0.3$ to $0.03$ and corresponding $\lambda_{bb}$ increases from $30$ to $100$. It is very much likely that the measurement of $P_{N}$ and its average value will help us to distinguish the NP effects coming through different versions of the THDM.

In Eq. (\ref{Transverse-polarization}), we can see that the transverse lepton polarization asymmetry $(P_T)$ is not only $m_{\ell}$ suppressed but it is also proportional to the imaginary part of the different combinations of the Wilson coefficients. Therefore, its value is expected to be too small to measure experimentally, therefore, we have not shown it graphically in the present study.

\section{Conclusion}

The experimental results on angular observables in the rare decay $B \to K^{*} \ell^+ \ell^-$ have shown some deviations from the SM predications \cite{LHCb, ATLAS, CMS} and these observables are investigated in detail in literature \cite{Virto, 1308}. It has been pointed out that in certain observables like $\mathcal{P}_5^{\prime}$, where the deviations from SM predictions are $2 - 3$$\sigma$, it is possible to accommodate certain NP and it will be interesting if one do such analysis in different versions of Two Higgs Doublet Model. However, the purpose here is to give an overview of the NP coming through the allowed parameteric space of the THDM on the forward-backward asymmetry and different lepton polarization asymmetries in $B \rightarrow K_{1} \ell^{+} \ell^{-}$ decays. We observed that
the forward-backward asymmetry and the different lepton polarization asymmetries show a clear signal of the THDM model of all the three types. However, the $CP$ violation asymmetry is only non-zero in the type III of the THDM and it is because of the presence of new phase $\delta$ and it will be discussed in a separate study \cite{IAJ}. Therefore, the precise measurement of this asymmetry along with the one calculated here will help us to get the constrains on the phase $\delta$ as well as other parameters of the THDM.

To sum up, the more data to be available from LHCb and the future super B-factories  will provide a powerful testing ground for the SM and also put some constraints on the Two Higgs doublet model parameter space.

\section*{Acknowledgments}
The authors M. J. A. would like to thank the financial support by the Quaid-i-Azam University from the University Research fund. I. A. would like to thanks the support by the S÷ao Paulo Research Foundation (FAPESP) under grant no. 2013/23177-3.

\end{document}